\documentclass[12pt]{article}
\usepackage{amsmath}
\usepackage{amssymb}
\usepackage{epsf}
\usepackage{cite}
\newcommand{\ads}[1]{{\rm AdS}_{#1}}

\def\Gh{\widehat{G}}
\def\muh{{\widehat{\mu}}}

\def\It{{\widetilde{I}}}
\def\jt{{\widetilde{j}}}
\def\Jt{{\widetilde{J}}}
\def\Nt{{\widetilde{N}}}
\def\st{{\widetilde{s}}}

\def\alphat{{\widetilde{\alpha}}}
\def\betat {{\widetilde{\beta}}}
\def\gammat{{\widetilde{\gamma}}}
\def\sigmat{{\widetilde{\sigma}}}
\def\Psit{{\widetilde{\Psi}}}

\def\psit{{\widetilde{\psi}}}
\def\Phit{{\widetilde{\Phi}}}
\def\phit{{\widetilde{\phi}}}

\def\tb{{\bar{t}}}
\def\wb{{\overline{w}}}
\def\zb{{\overline{z}}}
\def\partialb{{\bar{\partial}}}

\def\CA{{\cal A}}

\def\CM{{\cal M}}
\def\CN{{\cal N}}
\def\CO{{\cal O}}

\def\half{{\frac{1}{2}}}

\def\bracket#1{{\langle #1 \rangle}}
\def\Bracket#1{{\left\langle #1 \right\rangle}}
\def\ket#1{{\left| #1 \right\rangle}}
\def\p{\partial}
\def\pb{\bar{\partial}}

\def\Fv{{\bf F}}
\def\xv{{\bf x}}

\numberwithin{equation}{section} % make eq labels (sec.num)
\begin{document}

%%
%% title page (a la harvmac)
%%

\begin{flushright}
\vspace*{-1cm}
{\tt hep-th/0508110}\\
UPR-1128-T\\
UCLA-TEP-05-26\\
CALT-68-2573

%{\tt \jobname.tex}\\
%\today~~\now
\end{flushright}

\begin{center}
%\vspace*{1.5cm}
\vspace*{0.5cm}
%{\LARGE How black holes emerge from pure states}
%{\LARGE Typical half-BPS states of the D1-D5 system and their effective description as a black hole}
%{\Large The massless BTZ black hole as the effective geometry  }\\
%{\Large of the D1-D5 system}
%{\Large How the massles BTZ black hole emerges from the D1-D5 microstates}
{\LARGE Massless black holes and black rings\\[1ex] as effective geometries of the D1-D5 system}
\\
%\vspace*{.2cm}
%\centerline{\LARGE and the second line}\\
%\vspace*{1.7cm}

\vspace*{1cm}

Vijay Balasubramanian$^1$, Per Kraus$^2$, and Masaki Shigemori$^3$\\
\vspace*{0.7cm}
$^1$ David Rittenhouse Laboratories, University of Pennsylvania,\\
 Philadelphia, PA 19104, USA\\[1ex]
$^2$ Department of Physics and Astronomy, UCLA\\ Los Angeles, CA 90095-1547, USA\\[1ex]
$^3$ California Institute of Technology 452-48, Pasadena, CA 91125, USA\\
\vspace*{0.5cm}
{\tt vijay@physics.upenn.edu},  \quad
{\tt pkraus@physics.ucla.edu},\\
{\tt {s}{h{}}{i{}}{{}}g{}e@{}t{he{}}or{{}}y{}{.{}calt{}e{}ch.edu}} % aviod spam
\end{center}
%\vspace*{1.5cm}

\vspace*{0.7cm}

%%
%% abstract
%%

\baselineskip=18pt % a la harvmac

%We study the emergence of effective geometries describing complex
%underlying microstates of spacetime.

We compute correlation functions in the AdS$_3$/CFT$_2$
correspondence to study the emergence of effective spacetime
geometries describing complex underlying microstates. The basic
argument is that almost all microstates of fixed charges lie close
to certain ``typical'' configurations.  These give a universal
response to generic probes, which is captured by an emergent
geometry. The details of the microstates can only be observed by
atypical probes. We compute two point functions in typical ground
states of the Ramond sector of the D1-D5 CFT, and compare with
bulk two-point functions computed in asymptotically AdS$_3$
geometries. For large central charge (which leads to a good
semiclassical limit), and sufficiently small time separation, a
typical Ramond ground state of vanishing $R$-charge has the $M=0$
BTZ black hole as its effective description.  At large time
separation this effective description breaks down. The CFT correlators
we compute take over, and give
a response whose details depend on the microstate. We also discuss
typical states with nonzero $R$-charge, and argue that the
effective geometry should be a singular black ring.  Our results
support the argument that a black hole geometry should be
understood as an effective coarse-grained description that
accurately describes the results of certain typical measurements,
but breaks down in general.

%We study how effective bulk geometries emerge from CFT
%microstates in the AdS/CFT correspondence.  Working in the context
%of the D1-D5 system on $T^4$, we compute various two-point
%correlation functions evaluated in particular Ramond-Ramond ground
%states of the CFT\@.  These are compared with bulk two-point
%functions computed in asymptotically AdS$_3$ geometries. For $N_1
%N_5 \gg 1$, and for sufficiently small time separation, we find
%that if we choose a typical state in the CFT then we recover the
%bulk geometry of the $M=0$ BTZ black hole. At large time
%separation the effective bulk description breaks down and we
%observe the quasi-periodic behavior expected of a finite entropy
%system.  We also discuss typical states with nonzero $R$-charge, and
%argue that the effective geometry should be that of a singular black
%ring.  Our results support the argument that a black hole geometry
%should be understood as an effective coarse-grained description
%that accurately describes the results of certain typical
%measurements, but breaks down in general.

%%
%% main text
%%

\newpage
\setcounter{page}{1} % don't number title page

%\tableofcontents

\section{Introduction}

The AdS/CFT correspondence
\cite{Maldacena:1997re,Witten:1998qj,Gubser:1998bc} has provided a
detailed connection between black holes and conformal field
theories.  Although it is sometimes said that this solves the
conceptual puzzles associated with black hole physics, in fact we
still don't understand the connection well enough to see
explicitly how all these puzzles are resolved.

In using the AdS/CFT correspondence in the context of black holes one
typically compares a thermal ensemble in the CFT to a semi-classical
black hole geometry in the bulk. In this way it is possible to compute
and compare quantities such as the entropy of the system and correlation
functions of fields/operators \cite{Maldacena:2001kr, Hemming:2002kd,
Kraus:2002iv, Levi:2003cx, Fidkowski:2003nf, Barbon:2003aq,
Balasubramanian:2004zu, Kleban:2004rx,Festuccia:2005pi}.  Recent work
\cite{Behrndt:1998eq,Ooguri:2004zv,Dabholkar:2004yr,Dabholkar:2004dq,
Dabholkar:2005by,Hubeny:2004ji,Sen:2005pu,Kraus:2005vz,Iizuka:2005uv}
has shown that in some cases one can do even better by extending this
relation to the regime where the bulk geometry receives large
corrections from higher derivative string and loop effects.

In the CFT it is manifest that the thermal ensemble corresponds to a
weighted collection of individual microstates. Instead of
considering such an ensemble, there is nothing to prevent one from
choosing a particular microstate and computing correlation functions
in that state.  On general grounds, if one is working at large $N$
then correlation functions of ``typical'' operators computed in a
``typical'' state will be approximated to excellent accuracy by the
same correlators computed in the thermal ensemble. This is just the
same as saying that realistic isolated systems, i.e.\ a large number
of  molecules in a sealed box, are in some particular quantum
mechanical microstate at a given time, yet can be accurately studied
by the methods of statistical mechanics and thermodynamics.

The most natural interpretation of the AdS/CFT correspondence is
that there is a one-to-one correspondence between bulk and boundary
states.  In particular, one expects this statement to hold even in
the range of parameters where black holes are allowed.  Precisely
what the bulk microstates should look like is unclear at this point
in time.  Mathur  \cite{Lunin:2001jy} has conjectured that these
microstates correspond to bulk geometries (in general these might be
classically singular or have large quantum fluctuations) without
horizons, and the evidence for this conjecture includes
\cite{Lunin:2001jy,Mathur:2003hj,Bena:2004wt,Bena:2005ay,
Jejjala:2005yu,Bena:2005va,Berglund:2005vb,Taylor:2005db}.
Just as in the CFT, one is led to believe that if one chooses a
typical such bulk state then with respect to typical measurements it
will look like the usual black hole geometry.

An example of an {\it atypical\/} measurement is one which extends over
a very long time interval.  As originally emphasized by Maldacena
\cite{Maldacena:2001kr}, at late times correlators computed in the
semi-classical black hole geometry decay to zero, while in the CFT they
exhibit a quasi-periodic behavior.  The key difference is that the
semi-classical black hole geometry has a continuous spectrum due to the
presence of the horizon, while the CFT has a discrete spectrum. This
distinction is insignificant for short times or at high energy, but
becomes important in the opposite regime. Further work
\cite{Barbon:2003aq,Kleban:2004rx} in the context of BTZ black holes~\cite{btz}
strongly indicates that also summing over the $SL(2,\mathbb Z)$ images of the
black hole (which includes global AdS$_3$) can prevent the correlators
from decaying to zero, but can't account for the
quasi-periodicity. Presumably, this is telling us that we should instead
be considering the actual microstate geometries dual to the individual
CFT microstates if we want to correctly account for detailed properties
such as the late time behavior of correlators. The analogy with
molecules in a box is again helpful: while a coarse-grained effective
description accurately describes most properties of the system, in order
to recover the quasi-periodicity of late time correlators one needs to
return to the fundamental molecular description.

Recently, some of these issues have been discussed in the context of the
$\ads{5}/{\rm CFT}_4$ correspondence for half-BPS states
\cite{Corley:2001zk, Berenstein:2004kk, Lin:2004nb}.  It was argued in
\cite{Balasubramanian:2005kk} that typical large-charge half-BPS
microstates that are incipient black holes have a spacetime description
as a quantum ``foam'', the precise details of which are almost invisible
to almost all probes.  This gave rise to effective singular descriptions
of underlying smooth quantum states \cite{Balasubramanian:2005kk}.
(Other perspectives on these issues have appeared in \cite{others}.)  In
the present paper we will study the D1-D5 system on $T^4$, since this
provides the simplest link between black holes and CFT\@.  We will also
set to zero the momentum $P$, so that we just work with the
Ramond-Ramond ground states of the system.  This example is of interest
for several reasons. The system has a large ground state degeneracy
corresponding to an entropy $S= 2\pi \sqrt{2}\sqrt{N_1 N_5}$, and so
should have some of the properties of a black hole.\footnote{In the case
of the D1-D5 system on K3 it has been shown that higher curvature terms
indeed lead to an event horizon whose entropy coincides with that in the
CFT \cite{Dabholkar:2004yr,Dabholkar:2004dq}.  Whether this also happens
in the $T^4$ case is unclear
\cite{Behrndt:1998eq,Ooguri:2004zv,Dabholkar:2004yr,Dabholkar:2004dq,
Dabholkar:2005by,Hubeny:2004ji,Sen:2005pu,Kraus:2005vz,Iizuka:2005uv}.}
A large class of microstate geometries for this system is known.  They
correspond to configurations in which the D1 and D5 branes expand into a
Kaluza--Klein monopole supertube \cite{Lunin:2002iz}.  The shape of the
supertube encodes the details of the microstate.  For example, the
maximally $R$-charged microstate corresponds to a circular supertube.
On the other hand, a typical state corresponds to the supertube taking a
complicated random walk shape localized near the origin.  This leads to
a strongly curved supergravity solution whose existence is inferred via
extrapolation from the weakly curved geometries described by smooth
curves of large size.

We will compute correlation functions of certain operators in typical
states of the D1-D5 system.  As we have already discussed, at large
$N=N_1 N_5$ the expectation is that these correlators should coincide
with bulk correlators computed in some effective geometry.  This
effective geometry is analogous to the black hole geometry in the case
of the D1-D5-P system.  As we will see, the effective geometry that
emerges depends on the $R$-charge of the underlying state.  If the
$R$-charge vanishes, the emergent geometry is the massless BTZ
\cite{btz} black hole; or equivalently, AdS$_3$ in Poincar\'{e}
coordinates with a spatial direction periodically identified. This is
often referred to as the ``naive'' geometry representing the RR-ground
states.  It can be obtained by contracting the KK-monopole supertube to
zero size.  No individual microstate corresponds to this geometry;
rather, in the large $N$ limit, this geometry encodes the universal
response of generic finite time correlation functions in the underlying
microstate.  This effective geometry exhibits a continuous spectrum just
like the black hole, and so bulk correlators decay to zero at late
times.  But in this case we can also show that the exact late time
correlation functions show quasi-periodic behavior demonstrating that
the effective description breaks down at large times, and should be
replaced by the exact microstate geometries.

We would like to emphasize that our approach based on computing
correlation functions allows us to {\em derive\/} the effective
geometries corresponding to CFT states, rather than assuming the (highly
plausible!)  map between states and KK-monopole supertube profiles.  A
correlation function based approach is also necessary if one wishes to
make statements about the geometry at the string or Planck scale, since
the existing map between states and geometries is only valid at the
level of two-derivative supergravity.

We also consider typical states of nonzero $R$-charge, and find some new
features.  For sufficiently large charge, the CFT undergoes a form of
Bose--Einstein condensation.  The state effectively splits into two
components, one carrying the $R$-charge but no entropy, and the other
carrying no $R$-charge and all the entropy.  The correlation functions
we compute then become a sum of two terms with contributions from each
of the two components.  The result looks effectively like a
superposition of correlators computed in the massless BTZ space and in
the maximal $R$-charge Ramond vacuum, namely globals $\ads{3}$ with a
Wilson line \cite{Balasubramanian:2000rt,Maldacena:2000dr,
Lunin:2002iz}.  On the other hand, we are able to derive a prediction
for the effective bulk geometry with non-zero $R$-charge by explicitly
constructing the typical microstate geometry and coarse-graining it.  In
effect, this amounts to adding fluctuations on top of the circular
supertube solution and then averaging over these fluctuations in the
large $N$ limit.  The predicted effective geometry for non-zero
$R$-charge turns out to be a singular black ring solution.  It would
interesting to establish the connection between this prediction and the
``superposition'' of geometries derived from the CFT correlators.

The ultimate goal of this sort of investigation is to see a
macroscopic semi-classical black hole geometry emerging from
correlators computed in a typical state of a large $N$ CFT\@. This
requires knowing how the presence of a black hole manifests itself
in terms of correlators.  Let us note two criteria.  First, the
correlators should correspond to a well-defined classical
geometry, rather than a strongly fluctuating superposition.
Second, correlators should fall to zero at late times (in the
large $N$ limit), reflecting the presence of a horizon.  Here we
observe that the effective geometry emerging from our computations
satisfies these two properties, and so we can claim to be seeing
some black hole-like properties.  For a large black hole one would
like to do better and  reproduce the most important property of
all --- that of complete absorption of high energy particles
impinging on the horizon, but this goes beyond what we can do
here.

The remainder of this paper is organized as follows.  In section
\ref{sec:d1d5} we review the physics of the D1-D5 system, summarizing
he relevant details of the D1-D5 CFT, the map between Ramond ground
states and microstate geometries, and the computation of 2-point
correlation functions for massless scalars in the black hole spacetimes.
A more extensive review explaining details appears in Appendix
\ref{app:cft}.  In section \ref{sec:typ}, we construct the typical
Ramond ground states of the D1-D5 CFT that have fixed $R$-charges.  In
section \ref{sec:effgeo} we derive the effective geometries describing
finite-time correlation functions computed in these microstates.  The
basic technique is to compute and analyze the two-point correlator in
the typical states constructed in section \ref{sec:typ}.  In section
\ref{sec:disc} we conclude.  Appendix \ref{app:J=/=0} gives additional
details about typical states with nonzero $R$-charge.

%\section{D1-D5 bulk geometries and correlators}
\section{The D1-D5 system and its geometric dual}
\label{sec:d1d5}

\subsection{The D1-D5 CFT and its Ramond sector ground states}
\label{subsec:d1d5andR}

Consider type IIB string theory on $S^1 \times T^4$ with $N_1$ D1-branes
and $N_5$ D5-branes. The D1-branes are wound on $S^1$ and the D5-branes
are wrapped on $S^1\times T^4$.  At low energies, the worldvolume
dynamics of the branes is given by an $\CN = (4,4)$ supersymmetric sigma
model whose target space is the symmetric product $\CM_0=(T^4)^N/S_N$,
where $S_N$ is the permutation group of order $N$
\cite{Strominger:1996sh, deBoer:1998ip, Seiberg:1999xz, Larsen:1999uk}.
Here we set
\begin{align}
 N&\equiv N_1 N_5.
\end{align}
More precisely, $\CM_0$ is the so-called orbifold point in a family of
CFTs which are regained by turning on certain marginal deformations of
the sigma model on $\CM_0$.  At the orbifold point the CFT becomes free.
The D1-D5 CFT is dual to Type IIB string theory on $\ads{3} \times S^3
\times T^4$, which is the near-horizon limit of the D1-D5 brane system.
The AdS$_3$ length scale is given by $\ell \sim N^{1/4}$.  To have a
large, weakly coupled, $\ads{3}$ space, $N$ must be large and the CFT
must be deformed far from the orbifold point.  This situation is
familiar in the AdS$_5$/SYM$_4$ duality, where the SYM theory becomes
free at a special point ($g_{\rm YM}=0$) in the moduli space, but in
order for it to correspond to a semiclassical gravity one has to turn on
the coupling $g_{\rm YM}$.  The orbifold point is the analogue of the
free SYM\@.  In the following, we will consider the orbifold point of
the D1-D5 CFT, so one should bear in mind that exact agreement with
computations in supergravity is not expected in all cases, although some
protected BPS quantities can be computed exactly.

According to the AdS/CFT correspondence, every pure state of the D1-D5
CFT is dual to a pure state of string theory in $\ads{3} \times S^3
\times T^4$.  Here, we are interested in understanding how black holes
emerge as the effective spacetime description of underlying pure states
in gravity.  The only black holes in $\ads{3}$ gravity with standard
boundary conditions are the BTZ solutions \cite{btz}.  The
supersymmetric versions of these spacetimes have periodic boundary
conditions for fermions around the asymptotic circle in the $\ads{3}$
geometry and thus appear in the Ramond sector of the theory.
Furthermore, the lightest of the black holes, the BPS massless solution,
has the quantum numbers of a ground state in the Ramond sector of the
dual CFT \cite{strominger}.  For these reasons, we will concentrate on
the Ramond ground states of the D1-D5 CFT henceforth.  Powerful
techniques to study these ground states are available at the orbifold
point of the CFT\@.

% The simplest black hole that we can study in $\ads{3}$  is the half-BPS massless BTZ black hole, which appears in the Ramond sector of the theory, with periodic boundary conditions for fermions around the asymptotic circle in the $\ads{3}$ geometry \todo{cite}.     In the dual D1-D5 CFT, the $M=0$ BTZ black hole has the quantum numbers of a ground state in the Ramond sector.
%Powerful techniques to study such states are available at the orbifold point of the CFT.
%  As we said above, to describe a large spacetime, the CFT should be deformed away from this special point, but we will not do this here because we are primarily interested in certain structural properties of half-BPS states which should be preserved even at finite coupling.

The D1-D5 CFT and the construction of the Ramond ground states is
reviewed in detail in Appendix \ref{app:cft}\@.  For the moment, the
following facts are sufficient.  At the orbifold point we are dealing
with an ${\cal N} = (4,4)$ SCFT on the target space ${\cal M}_0 =
(T^4)^N/S_N$.
%This theory has an $SU(2)_R \times \widetilde{SU(2)}_R$
%$R$-symmetry, which is realized in the dual string theory as rotations
%of the $S^3$ factor in $\ads{3} \times S^3 \times T^4$.
This theory has an $SU(2)_R \times \widetilde{SU(2)}_R$ $R$-symmetry,
which originates from the $SO(4)$ rotational symmetry transverse to the
D1-D5 worldvolume.
There is
another global $SU(2)_I \times \widetilde{SU(2)}_I$ which is broken by
the toroidal identifications in $T^4$, but can be used for classifying
states anyway.  We will label the charges under these symmetries as
\begin{equation}
(J^3_R, \tilde{J}^3_R) = (\tfrac{s}{2}, \tfrac{\st}{2})
 ~~~~ {\rm and}  ~~~~
 (I^3, \tilde{I}^3) = (\tfrac{\alpha}{2}, \tfrac{\alphat}{2})
\label{chargedef}
\end{equation}
with $s,\tilde{s},\alpha,\alphat = \pm 1$.  The CFT has a collection of
twist fields $\sigma_n$, which cyclically permute $n \leq N$ copies of
the CFT on a single $T^4$.  One can think of these operators as creating
winding sectors of the worldsheet that wind over the different copies of
the torus.  The product of twist operators is also a twist operator.
The elementary bosonic operators of twist $n$ carry either
$SU(2)_R\times\widetilde{SU(2)}_R$ or $\widetilde{SU(2)}_I$ charges
($\sigma_n^{s\tilde{s}}$ or $\sigma_n^{\alphat\betat}$), while the
elementary fermionic twist operators are charged under $SU(2)_R \times
\widetilde{SU(2)}_I$ or $\widetilde{SU(2)}_I\times\widetilde{SU(2)}_R$
($\tau_n^{s\alphat}$ or $\tau_n^{\alphat\st}$).  A general Ramond sector
ground state is constructed by multiplying together elementary bosonic
and fermionic twist operators to achieve a total twist of $N = N_1 N_5$:
\begin{equation}
\begin{split}
 \sigma&= \prod_{n,\mu} (\sigma_{n}^{\mu})^{N_{n\mu}} (\tau_{n}^\mu)^{N'_{n\mu}},
 \\
 \sum_{n,\mu}n (N_{n\mu}+N'_{n\mu})&=N, \qquad
 N_{n\mu}=0,1,2,\dots,\quad N'_{n\mu}=0,1 \, .
\end{split}\label{gen_twist2}
\end{equation}
Here $\sigma_n^\mu$ and $\tau_n^\mu$, are the constituent elementary
twist operators, and $\mu = 1 \cdots 8$ labels their possible
polarizations ($\mu=(s,\st),(\alphat,\betat)$ for bosons, and $\mu =
(s,\alphat), (\alphat,\st)$ for fermions).  Appendix \ref{app:cft} gives
a detailed description of the construction of the twist operators and
computations using them.  For our immediate purposes, the relevant point
is that the integers
\begin{equation}
\{ N_{n\mu}, N^\prime_{n\mu} \}
\label{Rstateintegers}
\end{equation}
uniquely specify a Ramond ground  state.

\subsection{Map to the FP system and microstate geometries}

It has been proposed that each Ramond ground state of the D1-D5
has a corresponding exact spacetime geometry without horizons
\cite{Lunin:2001jy}.  The construction of these geometries was
carried out by first $U$-dualizing the D1-D5 system to the FP system
in type II, where an F1 string is wound $N_5$ times along $S^1$
and carries $N_1$ units of momentum in the $S^1$ direction
\cite{Lunin:2001fv}.  If the right-moving oscillation number $N_R$
vanishes, $N_R=0$, then this configuration is BPS\@. Such states
can be written as
\begin{equation}
 \label{FPstates1}
\begin{split}
 & \prod_{n,\mu} (\alpha_{-n}^{\mu})^{N_{n\mu}}
  (\psi_{-n}^\mu)^{N_{n\mu}'}\ket{N_1,N_5},
 \\
 N_L=\sum_{n,\mu}n(N_{n\mu}+N_{n\mu}')&=N_1N_5=N, \qquad
 N_{n\mu}=0,1,2,\dots,\quad N_{n\mu}'=0,1.
\end{split}
\end{equation}
Here $\alpha_{-n}^\mu$ and $\psi_{-n}^\mu$ are left-moving bosonic and
fermionic oscillators, respectively.  The polarization $\mu$ runs over
eight transverse directions.  $\ket{N_1,N_5}$ is the F1 string state
with momentum $N_1$ and winding number $N_5$, and with no oscillators
excited (this state itself is not physical).  The second line displays
the Virasoro constraint on the left-moving oscillation number $N_L$.

Following \cite{Lunin:2001jy, Lunin:2002iz}, the $U$-duality map between
the states (\ref{FPstates1}) of the FP system and the Ramond ground
states (\ref{gen_twist2}) of the D1-D5 system is given by:
\begin{equation}
\sigma_n^\mu\leftrightarrow\alpha_{-n}^\mu
~~~~;~~~~
\tau_n^{\mu}\leftrightarrow\psi_{-n}^\mu \, .
\label{FPD1D5map}
\end{equation}
The set of integers (\ref{Rstateintegers}) defining a Ramond ground
state is precisely mapped into the set of integers defining an
excitation of the FP system.

The metric of the FP system is known for arbitrary classical profile
$x^\mu=F^\mu(v)$ of the F1 string by the chiral null model
\cite{cmp,dghw,Horowitz:1994rf,Tseytlin:1996yb}.  Here $\mu$ runs over
the eight transverse directions to the F1 worldsheet. $v=t-y$ is the
left-moving lightcone coordinate, reflecting the fact that there must be
only left-moving waves on the F1 string because of the BPS condition.
By $U$-dualizing back, Lunin and Mathur \cite{Lunin:2001jy} obtained the
metric of the D1-D5 system, when the classical profile $F^\mu(v)$ is
only in the noncompact $\mathbb R^4$ directions $\xv=x^i$,
$i=1,2,3,4$. Explicitly, the string frame metric of the D1-D5 system in
the decoupling limit is given by \cite{Lunin:2001jy, Lunin:2002iz}
\begin{equation}
 \begin{split}
  ds_{\rm string}^2&={1\over\sqrt{f_1f_5}}[-(dt-A)^2+(dy+B)^2]+\sqrt{f_1f_5}\,dx^i dx^i
  +\sqrt{f_1\over f_5}dz^a dz^a,\\
  e^{2\Phi}& ={f_1\over f_5},\qquad
  f_5 ={Q_5\over L}\int_0^L{dv\over |\xv-\Fv(v)|^2},\qquad
  f_1 ={Q_5\over L}\int_0^L{|\dot \Fv(v)|^2 dv\over |\xv-\Fv(v)|^2},\\
  A_i& =-{Q_5\over L}\int_0^L{\dot F_i(v) dv\over |\xv-\Fv(v)|^2},\qquad
  dB=-*_4 dA.
 \end{split}\label{LMmetric1}
\end{equation}
Here, $y$ and $z^a$ are $S^1$ and $T^4$ directions, respectively.  The
coordinate radius of $S^1$ is $R$, and the coordinate volume of $T^4$ is
$(2\pi)^4V_4$.  The length $L$ is related to $R$ by
\begin{align}
 L={2\pi N_5\over R},
\end{align}
where the D5-brane charge $Q_5$ is related to the D5 number $N_5$ by
$Q_5=g_s\alpha' N_5$.  The four arbitrary functions $\Fv(v)=F_i(v)$,
$0\le v\le L$ in \eqref{LMmetric1} parametrize the solution, and
correspond to the classical profile $\Fv_{\rm FP}(v)$ of the F1 string
in the FP duality frame by $\Fv(v)=\mu \Fv_{\rm FP}(v)$, $\mu=g_s
\alpha'{}^{3/2}/R\sqrt{V_4}$ \cite{Lunin:2001jy}.  The D1 charge is
given by
\begin{align}
 Q_1&={Q_5\over L}\int_0^L{|\dot \Fv(v)|^2 dv}.
 \label{Q12}
\end{align}
In this paper we will argue that typical probes of typical microstate
geometries will react as if the spacetime was simply an $M=0$ BTZ black
hole (\ref{naive}) below.

Using (\ref{FPD1D5map}) the Ramond ground states can be mapped onto
specific states of the FP system \eqref{FPstates1}. For states involving
only $\alpha^i_{-n}$ this in turn determines the classical profile
$F^i(v)$ of the F1 string, which can be substituted into
\eqref{LMmetric1} to give the proposed geometry corresponding to a
specific Ramond ground state. Details and examples are given in
\cite{Lunin:2001jy}.  For example, the special Ramond ground state
$[\sigma_{n}^{s\tilde{s}}]^{N/n}$ with $s=\tilde{s} = -1$, $1\le n\le N$
corresponds to the bulk geometry $(AdS_3\times S^3)/\mathbb Z_n \times
T^4$.  The 3-dimensional part of this geometry is the conical defect
described below.  For general states involving all bosonic oscillators
as well as fermionic oscillators more work is needed; see
\cite{Taylor:2005db} for results regarding the fermionic states.
%studied in subsection \ref{con_dfct}.

\subsection{Bulk geometries and correlators}

In this subsection we will review some geometries that show up as the
bulk geometries in the context of AdS/CFT for the D1-D5 system.  We will
also present the bulk 2-point functions of a massless minimally coupled
scalar in those geometries, and compare them at the end.  The asymptotic
$\ads{3}$ radius is given by $\ell\sim N^{1/4}$.

\subsubsection{Conical defect}
\label{con_dfct}

The Ramond ground state $[\sigma_{n}^{s\tilde{s}}]^{N/n}$ with
$s=\tilde{s} = -1$, $1\le n\le N$ corresponds in the bulk to the conical
defect geometry \cite{Balasubramanian:2000rt,Maldacena:2000dr}:
\begin{align}
ds^2 = -\left({1 \over n^2}+{r^2\over\ell^2}\right)dt^2
+ {dr^2 \over{1 \over n^2}+{r^2\over \ell^2}} + r^2 d\phi^2~. \label{condef}
\end{align}
Here $n$ is an integer in the range $1 \leq n \leq N$.  The angular
identification is $(\phi,\psi) \cong (\phi+2\pi, \psi +{2 \pi \over
n})$, where $\psi$ is angle on the  $S^3$ factor that we have
suppressed.  The special case $n=1$ yields AdS$_3$ in global
coordinates.

Any state can be probed by computing the correlation functions of
operators in that state.  The simplest correlator that one could
compute, the 2-point function, is related to a 4-point function computed
in the vacuum.  According to the AdS/CFT correspondence, the 2-point
function in the state $[\sigma_{n}^{s\st}]^{N/n}$, $s=\st=-1$ can be
obtained from AdS space by computing the bulk-boundary propagator of the
spacetime field that is dual to the CFT probe and then taking the bulk
point to the boundary.  Let us consider a CFT probe that is dual to a
massless scalar field in $\ads{3}$.  The conical defect propagator for
this field is obtained from the AdS propagator by summing over the
images that define the conical defect.  By translation invariance we can
take one of the boundary points to be at $t=\phi=0$. We then obtain the
result
\begin{align}
 \sum_{k=0}^{n-1}
 {1\over \Bigl(2n\sin{w-2\pi k\over 2n}\Bigr)^2\Bigl(2n\sin{\wb-2\pi k\over 2n}\Bigr)^2}
 =
 {1\over 16n^2\sin^2{w-\wb\over 2n}}
 \left[
 {1\over \sin^2{w\over 2}}+{1\over \sin^2{\wb\over 2}}
 -{2\sin{w-\wb\over 2}\over n\tan{w-\wb\over 2n}\sin{w\over 2}\sin{\wb\over 2}}
 \right]
 \label{conicalG}
\end{align}
where
\begin{align}
w =  \phi-{t \over \ell}, \quad \wb = \phi+{t \over \ell}~.
\end{align}
The summation was done by a standard contour integration method.

%\begin{align} G &= {1 \over 16n^4}\sum_{k=0}^{n-1}\left[ {1 \over \sin
%\left( {\ell^{-1} t -(\phi+ 2 \pi k ) \over 2n}\right) \sin \left(
%{\ell^{-1} t +(\phi+2\pi k) ) \over 2n}\right)}\right]^2 \cr & = {1
%\over 8 n^2 \sin^2 \left({t \over n \ell}\right)\sin^2 \left({t
%\over 2 \ell}\right)}\left[ 1+ {\sin \left({t \over \ell}\right)
%\over n \tan \left({t \over n \ell}\right)}\right]~.
%\end{align}
%

\subsubsection{Naive geometry}
\label{subsec:naive_geo}

Consider taking the $n\rightarrow \infty$ limit of the conical
defect geometries:
\begin{align} ds^2 = -{r^2\over \ell^2} dt^2 + {\ell^2 \over r^2}dr^2 + r^2
d\phi^2~. \label{naive}
\end{align}
This is the same as AdS$_3$ in Poincar\'{e} coordinates  with a
periodically identified spatial direction.  This geometry does not
actually correspond to any CFT microstate since it has $n>N$.
Instead, we will see that this geometry emerges as an effective
description of the typical Ramond ground state at  large $N$.  As
before we compute the boundary 2-point function for a massless
scalar field, and find that
\begin{align} \sum_{k=-\infty}^{\infty}
 {1 \over (w-2\pi k)^{2} (\wb-2\pi k)^{2}}={1 \over 4 (w-\wb)^2} \left[{1\over \sin^2{w\over 2}}+{1\over \sin^2{\wb\over 2}}
 -{4\sin{w-\wb\over 2}\over (w-\wb) \sin{w\over 2}\sin{\wb\over 2}}
 \right] ~.
 \label{naiveG}
\end{align}

\subsubsection{Non-rotating BTZ}

The above naive geometry is in fact the massless limit  of the BTZ
black holes of $\ads{3}$.  To see this recall that the non-rotating
BTZ black holes has a metric \cite{btz}
\begin{align} ds^2 = -{r^2- r_+^2 \over \ell^2} dt^2 + {\ell^2 \over
r^2- r_+^2 }dr^2+r^2 d\phi^2~.
\end{align}
If we take $r_+=0$, which is the $M=0$ BTZ black hole, we get back
the naive geometry (\ref{naive}).  The boundary 2-point function
is \cite{esko}
\begin{align}  {16 r_+^4 \over 4 \ell^4}\sum_{k=-\infty}^\infty \left[{1
\over \sinh \left({r_+ \over 2 \ell}(w+2\pi k)\right) \sinh
\left({r_+ \over 2 \ell}(\wb+2\pi k)\right) }\right]^2~.
\label{btzcorr}
\end{align}
We have not succeeded in doing the summation in closed form. But we
can use contour integration to rewrite the sum in a way which makes
the large time behavior manifest.  For simplicity set $\phi=0$. %%
%\begin{align} \sum_{n=-\infty}^\infty f(n) = - \sum \left( {\rm
%residues~of~} \pi f(z) \cot \pi z \right)
%\end{align}
%%
Then one can rewrite (\ref{btzcorr}) as
\begin{align} {r_+^2 \over 8 \ell^2} {1 \over \sinh^2 \left({r_+ t
\over \ell^2}\right)} \Biggl\{ \sum_{m=-\infty}^\infty & \left[
 { 1 \over \sin^2 \left({t \over 2 \ell}+i\pi{\ell\over r_+}m\right)}
  +{2r_+\over \ell}
   {1\over \tan\left({t \over 2 \ell}+i\pi{\ell\over r_+}m\right)
   \tanh\left({r_+ t \over \ell^2}\right)}
\right]\cr & +{2r_+\over \pi\ell} \left[ {\left({r_+ t\over
\ell^2}\right) \over \tanh\left({r_+ t\over \ell^2}\right)}-1\right]
\Biggr\}~.
\end{align}
To simplify further, consider the case of a small black hole, $r_+
\ll \ell$.  In this case we can truncate to just the $m=0$ term and
obtain:
\begin{align} {r_+^2 \over 8 \ell^2} {1 \over \sinh^2 \left({r_+ t
\over \ell^2}\right)}\Biggl\{{ 1 \over \sin^2 \left({t \over 2
\ell}\right)} +{2r_+ \over \ell} {1\over \tan\left({t \over 2
\ell}\right)
   \tanh\left({r_+ t \over \ell^2}\right)}  +{2r_+\over \pi\ell}
\left[ {\left({r_+ t\over \ell^2}\right) \over \tanh\left({r_+
t\over \ell^2}\right)}-1\right]\Biggr\}~.
\end{align}

\subsubsection{Comparison}

% This is a
%reflection of the fact that the conical defect geometries represent
%individual microstates of the CFT, while the naive and BTZ
%geometries are just effective descriptions.

Notice that the three geometries described above look the same outside  a core region.  As we'll
review later, the $n$ appearing in the conical defect geometry has a
typical size
\begin{align} n_{\rm typ} \sim N^{1/2} \sim \ell^2~.
\end{align}
The typical conical defect geometry thus approaches the naive
geometry for $r \gg  \ell^{-1}$.   Consider then the BTZ geometry
with $r_+ = \ell^{-1}$, so that it has the same characteristic size
as the typical conical defect.  The Bekenstein-Hawking entropy of
this black hole is then
\begin{align} S \sim A \sim \ell^3 r_+ \sim \ell^2 \sim N^{1/2}~,
\end{align}
where the factor of $\ell^3$ came from integration over the $S^3$
that we have suppressed.  $S \sim N^{1/2}$ is indeed  the correct
ground state entropy of the D1-D5 system. This is an example of the
stretched horizon idea advocated in \cite{Lunin:2001jy}.

The most obvious difference between the 2-point functions computed above
is that the conical defect result is periodic in time, with a period
$\Delta t = 2\pi n \ell$, while the naive geometry and the BTZ black
hole results decay to zero at large time.  Usually, this sort of decay
is associated with the presence of a horizon, with the information loss
problem arising because the decay winds up implying a failure of
unitarity \cite{Maldacena:2001kr}. Later in this paper we will show that
the decay is the correct universal description of the typical two-point
function in a typical state, but that its persistence to late times is
an artifact of ignoring the precise quantum mechanical details of the
individual microstates of a black hole.  Our computations will be for
the $M=0$ BTZ black hole which will turn out to be the effective
coarse-grained description of the typical Ramond ground state of the
dual CFT\@.  To this end, we now turn to characterizing the structure of
these states.

\section{Typical states}
\label{sec:typ}

\subsection{Statistics and typical states}

As described in subsection \ref{subsec:d1d5andR}, each ground state in
the Ramond sector of the D1-D5 CFT is characterized by a set of integers
$\{ N_{n\mu}, N^\prime_{n\mu} \}$ specifying the distribution of
constituent bosonic and fermionic twists (\ref{gen_twist2}).  When the
total twist length $N=\sum_{n,\mu} n (N_{n\mu}+N'_{n\mu})$ is very
large, there are a macroscopic number ($\sim e^{2\sqrt{2}\pi\sqrt{N}}$)
of Ramond ground states.  In such a situation, most of those
$e^{2\sqrt{2}\pi\sqrt{N}}$ microstates will have a twist distribution
$\{ N_{n\mu}, N^\prime_{n\mu} \}$ that lies very close to a certain
``typical'' distribution.  In the large $N$ limit, the difference among
individual distributions is small.  Roughly, statistical mechanics says
that $\langle(\Delta N_{n\mu})^2\rangle\sim N_{n\mu}$, thus
${\langle(\Delta N_{n\mu})^2\rangle^{1/2}\over N_{n\mu}}\sim
(N_{n\mu})^{-1/2}\to 0$ as $N_{n\mu }\to\infty$.  Thus, although
correlation functions computed in individual microstates depend on the
precise form of the microstate distribution $\{N_{n\mu},N_{n\mu}'\}$,
for almost all microstates the generic responses should deviate by small
amounts from the results for the typical state.  In the next section,
this will be the basis for the emergence of an effective black hole
description of typical Ramond ground states.  A similar analysis of the
typical states was carried out for the $\ads{5}/{\rm SYM}_4$ duality in
\cite{Balasubramanian:2005kk}.

In this section our goal is to characterize the typical distribution of
twists and the size of fluctuations around it within the ensemble of
Ramond ground states.  Ideally, we carry out a microcanonical analysis
by studying all partitions of integers of integers (\ref{gen_twist2})
that lead to a total twist of $N$.  However, it is easier to carry out a
canonical analysis by including states with arbitrary total twist into
the ensemble, while fixing the average total twist to be $N$ via an
effective temperature $T$.  The relative error incurred by the canonical
approach compared to the exact microcanonical analysis vanishes in the
large $N$ limit.  Large $N$ will correspond to large temperature $T \gg
1$, or equivalently, small $\beta = 1/T \ll 1$.  Since the constituent
twist operators in (\ref{gen_twist2}) carry an $R$-charge, we can study
the structure of Ramond ground states restricted to carry some fixed
$R$-charge. In particular, in terms of the $SU(2)_R \times
\widetilde{SU(2)}_R$ charges in \eqref{chargedef}, let us define:
\begin{align}
 J &=-J_R^3-\Jt_R^3,\qquad
 \tilde{J}=J_R^3-\Jt_R^3 \, .
\end{align}
With these definitions, the $R$-charges $J$ and $\tilde{J}$ correspond
in the bulk geometry to orthogonal angular momenta in the $\mathbb R^4$
perpendicular to the D1-D5 worldvolume.\footnote{If we let $\mathbb R^4$
coordinates be $x^{1,2,3,4}$, the two angular momenta in question are
$J_{12}$ and $J_{34}$.  See Appendix \ref{app:cft}.1 for details.}  We
will consider the structure of Ramond ground states with $\tilde{J} = 0$
and different values of $J$.

\subsection{Typical twist distribution with $J=\Jt=0$}

Let us first consider the ensemble of all the Ramond ground states
\eqref{gen_twist2} with equal statistical weight.  Some of the states in
this ensemble will have a non-vanishing $R$-charge.  However, because
the polarizations $\mu$ of twist operators $\sigma_n^\mu,\tau_n^\mu$ are
weighted equally, on average the states will have $J=\tilde{J} = 0$.
Indeed, as we will see, there are so many more states with $J=0$ than $J
\neq 0$ that summing over all states only incurs a small error in
studying the properties of $J=0$ states.

As described in (\ref{gen_twist2}) and Appendix \ref{app:cft}, we have 8
bosonic twist operators $\sigma_n^\mu$ and 8 fermionic twist operators
$\tau_n^\mu$ which are all independent.  So the canonical partition
function is
\begin{align}
 Z(\beta)&
 ={\rm Tr}[e^{-\beta N}]
 =\prod_{n=1}^\infty {(1+q^n)^8\over (1-q^n)^8}
 = \left[{\vartheta_2(0|\tau)\over 2\eta(\tau)^3}\right]^4,
 \qquad
 q=e^{2\pi i \tau}=e^{-\beta}.
\end{align}
Using the modular property of the theta function,
\begin{align}
 Z(\beta)
 = \left[{\beta\over 4\pi} {\vartheta_4(0|-{1\over\tau})\over \eta(-{1\over\tau})^3}\right]^4
 \sim e^{2\pi^2/\beta} \qquad (\beta\ll 1).
\end{align}
The relation between ``energy'' $N$ and temperature $\beta$ is
\begin{align}
 N&=\Bracket{\sum_{n=1}^\infty\sum_{\mu} n(N_{n\mu} +N'_{n\mu})}
 =-{\partial \over \partial \beta}\ln Z(\beta)
 \simeq {2\pi^2\over \beta^2}.\label{N_beta}
\end{align}
Since all twist operators are independent, the average
distribution $\{N_{n\mu},N'_{n\mu}\}$ is given by the
Bose--Einstein and Fermi--Dirac distribution, respectively:
\begin{align}
 N_{n\mu}&={1\over e^{\beta n}-1},\qquad
 N'_{n\mu}={1\over e^{\beta n}+1},\qquad
 N_n=\sum_\mu (N_{n\mu}+N'_{n\mu})={8\over \sinh \beta n}.
 \label{typ_J=0}
\end{align}
For large $N$, the typical states of our ensemble have a
distribution almost identical to \eqref{typ_J=0}.  We will call
the distribution \eqref{typ_J=0} the ``representative''
distribution.

\subsection{Typical twist distribution  with $J\neq 0$ and $\tilde{J} = 0$}
\label{subsec:typ_J=/=0}

Now let us consider the typical state in the ensemble with  fixed $R$-charge $J \neq 0$.
%angular momentum $J_{12}=J\neq 0$. The $R$-charges $J_R^3,\Jt_R^3$
%are related to the angular momentum $J_{12},J_{34}$ by
%\begin{align}
% J_{12}&=-J_R^3-\Jt_R^3,\qquad
% J_{34}=J_R^3-\Jt_R^3,
%\end{align}
From the definitions in subsection \ref{sec:d1d5}.1, the twist operators
that carry nonzero $J$ are
\begin{align*}
 \sigma_n^{s\tilde{s}}:&~  J=-(s + \tilde{s})/2, &
% \sigma_n^{--}:&~  J=+1, &
% \sigma_n^{++}:&~  J=-1, &
 \tau_n^{s\alphat}:&~  J=-s/2, &
 \tau_n^{\alphat\st}:&~  J=-\st/2.
\end{align*}
Strictly speaking, we should consider the microcanonical ensemble
in which $N$ and total $J$ are fixed.  But again in the large $N$
limit we can equivalently consider the canonical ensemble in which
$N$ and $J$ are controlled by temperature $\beta$ and chemical
potential  $\mu$.

To construct the partition function it is convenient to use the map
(\ref{FPD1D5map}) between the Ramond ground states and the FP system.
Then we are equivalently constructing the ensemble left-moving
oscillations of the FP string as specified in (\ref{FPstates1}).  In the
FP language we consider an ensemble in which we have $N_B$ left-moving
bosons $\alpha_{-n}^i$ and $N_F$ left-moving fermions $\psi_{-n}^i$,
where $n=1,2,\dots$. Let the bosons and fermions carry the following
$R$-charge assignments:
\begin{align}
\begin{array}{ccl@{~~~~~~~~~~~~}ccl}
 n_B      &\rm bosons:& J=+1& n_F      &\rm fermions:& J=+1/2\\
 n_B      &\rm bosons:& J=-1& n_F      &\rm fermions:& J=-1/2\\
 N_B-2n_B &\rm bosons:& J=0 & N_F-2n_F &\rm fermions:& J=0 \\
\end{array}\label{spin_assign}
\end{align}
The case we are interested in, {\it i.e.\/}\ the D1-D5 system on $T^4$,
has $N_B=N_F=8$, $n_B=1$, $n_F=4$.  The D1-D5 system on K3 has $N_B=24$,
$n_B=1$, $N_F=n_F=0$, for which state counting was first studied in
\cite{Russo:1994ev} from the heterotic dual perspective.  More recently,
the microscopics of the K3 case was studied in \cite{Iizuka:2005uv}, and
the discussions below and in Appendix \ref{app:J=/=0} are generalization
of the one therein.

We can compute the entropy $S(N,J)$ for given level $N$ and
$R$-charge  $J$ by studying the partition function
\begin{align}
 Z(\beta,\mu)\notag
 =\sum_{N,J}d_{N,J}q^N z^J
 ={\rm Tr}[e^{-\beta(N-\mu J)}]
 =\prod_{n=1}^\infty
 {[(1+z^{1/2}q^n)(1+z^{-1/2}q^n)]^{n_F}(1+q^n)^{N_F-2n_F}
 \over
 [(1-zq^n)(1-z^{-1}q^n)]^{n_B}(1-q^n)^{N_B-2n_B} },
% \notag\\
% &=2^{n_B-{N_F\over 2}}q^{N_B-N_F \over
% 24}\eta(\tau)^{-N_B+3n_B-{N_F\over 2}}
% \left[{\vartheta_2({\nu\over 2}|\tau)\over\cos{\pi \nu\over 2}}\right]^{n_F}
% \left[{\sin\pi\nu\over \vartheta_1(\nu|\tau)}\right]^{n_B}
% \vartheta_2(0|\tau)^{{N_F\over 2}-n_F}.
\end{align}
where $q=e^{2\pi i \tau}=e^{-\beta}$,  $z=e^{2\pi i
\nu}=e^{\beta\mu}$. The entropy in the $N\to\infty$ limit can be
evaluated by thermodynamic approximation, as explained in Appendix
\ref{app:J=/=0}, and the result is
\begin{align}
 S=\log d_{N,J}& =
 2\pi\sqrt{{c\over 6}(N-|J|)}~.
 \label{flyg27May05}
\end{align}
%\footnote{Note
%that Eq.\ (2.16) of \cite{Russo:1994ev} is incorrect; the power of the
%first factor should be $-(D+1)/4$, not $-(D+3)/4$.}
%Note that this result does not depend on $n_F$; all spins are carried by
%bosons.  This is because the Pauli exclusion principle exacts a high
%price in $N$ when the fermions carry a macroscopic amount ($\sim N$) of
%angular momentum.

Now let us apply this to the D1-D5 system  on $T^4$, for which
$N_B=N_F=8$, $n_B=1$, $n_F=4$.  One sees from \eqref{flyg27May05}
that the only effect of $J\neq0$ is to replace $N$ with $\Nt\equiv
N-J$. Here we assumed $J>0$. This means that almost all states
({\it i.e.}, the states that are responsible for the entropy) in
the ensemble with level $N$ and $R$-charge $J$ are of the
form
\begin{align}
 (\alpha_{-1}^+{}^\dagger)^J
 \underbrace{\prod_{n=1}^\infty
 \left[\prod_{i}(\alpha_{-n}^{i})^{N_{ni}}
 (\psi_{-n}^{i})^{N'_{ni}}\right]
 \ket{0}}_
 {\begin{minipage}{28ex}\scriptsize
   states that are responsible for entropy of the ensemble
  with level $\Nt=N-J$ and no angular  momentum
  \end{minipage}}
 ,\label{hlro26May05}
\end{align}
where $\alpha_{-n}^+{}^\dagger$ is the creation operator of the boson
that carries $J=+1$.  Indeed, the entropy from the
``$\underbrace{.....}$'' part is
$2\pi\sqrt{c\Nt/6}=2\pi\sqrt{c(N-J)/6}$, which fully accounts for
\eqref{flyg27May05}.  Of course, besides $ (\alpha_{-1}^+{}^\dagger)^J$
there are other combinations of oscillators that can carry $R$-charge
$J$.  But any other combination will exact more price in $N$, and will
therefore lead to a subleading contribution to the entropy.  If $J<0$,
then the same argument goes through if we replace
$(\alpha_{-1}^+{}^\dagger)^J$ with $(\alpha_{-1}^-{}^\dagger)^{|J|}$.

\bigskip
Translating the above into the language of the D1-D5 system,  the
typical state of the D1-D5 system with $N\gg 1$ and $J\neq 0$
splits into the following two parts:
\begin{enumerate}
 \item $|J|$ strings of unit length: $(\sigma_1^{s\st})^{|J|}$, where
       $s=\st=-1$ for $J>0$ and $s=\st=+1$ for $J<0$.  We will call this
       part the ``Bose--Einstein (BE) condensate''.
 \item the typical state of the ensemble with
       $\sum_{n\mu}n(N_{n\mu}+N'_{n\mu})=\Nt=N-|J|$ and no $R$-charge.
\end{enumerate}
In other words, the typical  distribution $\{N_{n\mu},N_{n\mu}'\}$
of the ensemble with level $N$ and angular momentum $J$ can be
written as
\begin{align}
 N_{n\mu}=N_{n\mu}^{\rm (BEC)}+\Nt_{n\mu},\qquad
 N_{n\mu}'=\Nt_{n\mu}',
 \label{typ_J=/=0}
\end{align}
where the BE condensate part $N_{n\mu}^{\rm (BEC)}$ is given by
\begin{align}
 \begin{cases}
  N_{n=1,\,s=\st=-1}^{\rm (BEC)}=J,   & {\rm other} ~~ N_{n\mu}^{\rm (BEC)}=0 \qquad (J>0),\\[2ex]
  N_{n=1,\,s=\st=+1}^{\rm (BEC)}=|J|, & {\rm other} ~~ N_{n\mu}^{\rm (BEC)}=0 \qquad (J<0),
 \end{cases}
\end{align}
while the non-condensate part $\{\Nt_{n\mu},\Nt_{n\mu}'\}$  is
identical to the typical distribution \eqref{typ_J=0} for the
ensemble with level $\Nt=N-|J|$ and no $R$-charge.

Note that the entropy of the ensemble with $J=0$ is  the same as
that in the ensemble with $J$ unspecified.  This is a reflection
of the fact that there are exponentially more states with $J=0$
than with $J\neq 0$.

%\subsection{Fluctuations and atypicality}

\section{The effective geometry}
\label{sec:effgeo}

In the previous section we derived the distributions of
constituent twist operators in typical Ramond ground states with
$R$-charges $J=0$ and $J \neq 0$.  In the large $N$ limit, almost
all states with the given charges have twist distributions that
lie close to these typical distributions.  Here we will compute
two-point correlation functions in typical states and show that
generic correlators computed at finite time separations are
largely independent of the details of the microstate. Indeed, at
small time separations two-point correlators in the typical $J=0$
state give a universal response, as if the corresponding spacetime
geometry was a black hole.  By contrast, atypical correlators
whose two insertion points are separated by very long times give
responses with intricate variations that encode the detailed
microstate.

The fact that generic probes of typical states give essentially
universal responses governed by the statistics of the state's
microscopic constituents is reminiscent of the similar observation in
\cite{Balasubramanian:2005kk} for the case of AdS$_5$/SYM$_4$. There, it
was argued that the correlation function of a small ``probe'' operator
$\CA$ in a black hole background produced by a long operator $\CO$ is
determined by the matching of patterns of fields ($X,Y,Z,\overline X$,
etc.) in both operators, the distribution of which is governed largely
by statistics.  In the case of half-BPS states of the AdS$_5$/SYM$_4$
theory \cite{Corley:2001zk, Berenstein:2004kk, Lin:2004nb} it was
possible to use the Yang--Mills theory to argue for an effective
spacetime description of microstates in terms of a singular geometry
\cite{Balasubramanian:2005kk}.  In the D1-D5 case, we have already noted
that there exists a proposed map taking RR ground states into
geometries, and we argue below that this map will also give an effective
singular spacetime description to typical states.  However, to genuinely
prove these assertions it is necessary to compute correlation functions,
because it is through correlators that we can rigorously compare bulk
and boundary physics in the AdS/CFT correspondence. Here we will carry
out the analysis of deriving an effective geometry from correlation
functions and thereby infer the effective geometry corresponding to a
typical state.  Furthermore, the maps between states and geometries,
either based on LLM \cite{Lin:2004nb} or on the Lunin-Mathur geometries,
are only valid at the level of two-derivative supergravity.  To learn
anything about the geometry when this approximation breaks down it is
necessary to extract the spacetime physics from CFT correlation
functions (or augment the original map with higher derivative terms).
We will give an explicit example of this here.  At late times our
correlators probe the strongly curved region of the geometry where the
effective spacetime description breaks down. The CFT correlators
continue to be valid and give a result which depends on which particular
microstate one has chosen.  This shows how the CFT can be used to go
beyond the accuracy of the low energy spacetime description.

\subsection{Two-point functions of the D1-D5 CFT}

For simplicity, we will compute the 2-point  functions of non-twist
``probe'' operators $\CA$ in states created by general twist
operators.   $\CA$ can be written as a sum over copies of the CFT:
\begin{align}
 \CA&={1\over \sqrt{N}}\sum_{A=1}^N \CA_A\label{sum_CA_A2}
\end{align}
where $\CA_A$ is a {\em non-twist\/} operator that lives  in the
$A$-th copy.  For example, we can take
\begin{align}
 \CA_A=\partial X^a_A(z) \bar \partial X^b_A(\zb),
 \label{A_grvtn2}
\end{align}
which corresponds to a fluctuation of the metric in the internal
$T^4$ direction.  Although, such non-twist operators are only a
subset of the operators that correspond to spacetime excitations, we
will restrict ourselves to them because their correlation functions
are much easier to compute than those of twist operators, and
because they will be sufficient to  demonstrate that an effective
geometry emerges in the $N\to \infty$ limit.

Given a general Ramond ground state $\sigma$ (\ref{gen_twist2})  we
are interested in computing
\begin{equation}
\langle \sigma^\dagger \CA^\dagger \CA \sigma \rangle
\end{equation}
The key result, demonstrated in Appendix \ref{app:cft}, is that for non-twist
operators at the orbifold point in the CFT such correlation
functions decompose into independent contributions from the
constituent twists operators in (\ref{gen_twist2}).     Denoting the
constituents $\sigma_n^\mu, \tau_n^\mu$ collectively by
$\sigma^{\hat{\mu}}_n$ and $N_{n\mu},N^\prime_{n\mu}$ by
$N_{n\hat{\mu}}$, we write the Ramond ground states
(\ref{gen_twist2}) as
\begin{equation}
\sigma = \prod_{n,\hat{\mu}} (\sigma_n^{\hat{\mu}} )^{N_{n\hat{\mu}}} \, .
\end{equation}
Then, the desired correlation function decomposes as
\begin{equation}
\langle \sigma^\dagger \CA^\dagger \CA \sigma \rangle
= {1 \over N}
\sum_{n,\hat{\mu}} n N_{n\hat{\mu}}
\sum_{A=1}^n \langle [\sigma^{\hat{\mu}}_n]^\dagger
 \CA_A^\dagger \CA_1 \sigma_n^{\hat{\mu}} \rangle~.
\label{SAAS2}
\end{equation}
The problem then reduces to computing 4-point functions of the form
\begin{align}
 \Bracket{ [\sigma_{(1\cdots n)}^{\muh}(z=\infty)]^\dagger \CA_A(z_1)^\dagger
 \CA_B(z_2) \sigma_{(1\cdots n)}^\muh(z=0)}
 \equiv
 \bracket{ \CA_A(z_1)^\dagger
 \CA_B(z_2) }_{\sigma_{(1\cdots n)}^\muh},
 \label{4pt_func2}
\end{align}
where $1\le A,B\le n$, and  the equation  indicates that we are
equivalently computing the 2-point function of $\CA$ in the ground
state of twist sector $n$.   As we described, in the $n$th twist
sector the worldsheet is effectively $n$ times as long and therefore,
as shown in Appendix \ref{app:cft}, for bosonic operators
\begin{align}
 \bracket{\CA_A^\dagger(w_1) \CA_B(w_2)}_{\sigma_{(1\cdots n)}}=
 {C \over \left[2 n \sin\left({w \over 2n }\right)\right]^{2h}
 \left[2n \sin\left({\wb \over 2n}\right)\right]^{2\tilde h}},
 \label{corrAA_w2}
\end{align}
where
\begin{align}
 w &\equiv w_1-w_2,\qquad
 \wb \equiv \wb_1-\wb_2.
\end{align}
Here, the copy labels $A,B$ mean that $w_1$ and $w_2$ must be understood
as $w_1+2\pi (A-1)$ and $w_2+2\pi (B-1)$, respectively.  The analogous
computation for fermionic $\CA$ is given in Appendix
\ref{app:cft:fermicorr}.

\subsection{Example of typical state correlation function}
\label{subsec:typ_corr}

The correlation function of non-twist operators in the general
microstate \eqref{gen_twist2} can be computed by plugging
\eqref{corrAA_w2} and \eqref{ferm_corr_bldblk} into the general
formula \eqref{SAAS2}.  For example, for $\CA$ purely bosonic, we
substitute bosonic correlator \eqref{corrAA_w2} into \eqref{SAAS2}
to obtain
\begin{align}
 \bracket{\CA(w_1) \CA(w_2)}_{\Sigma}
 ={1\over N}\sum_{n} n N_n \sum_{k=0}^{n-1}
 {C \over \left[2 n \sin\left({w-2\pi k\over 2n}\right)\right]^{2h}
 \left[2n \sin\left({\wb-2\pi k \over 2n}\right)\right]^{2\tilde h}},
 \label{AASigma}
\end{align}
where
\begin{align}
 N_n\equiv \sum_\mu (N_{n\mu}+ N_{n\mu}').\label{mnmc14Jul05}
\end{align}
Here we took into account that the copy labels $A,B$ mean that $w$ in
\eqref{corrAA_w2} should be replaced by $w+2\pi(A-B)$.  The correlation
function for the typical state is obtained simply by plugging the
typical distribution \eqref{typ_J=0} or \eqref{typ_J=/=0} into
$\{N_{n\mu},N_{n\mu}'\}$ above.  For fermionic $\CA$, the correlation
function \eqref{4pt_func} depends on the spin $\mu$ and the expression
is more complicated, as we will see below.

As a simple example of a probe $\CA$, take the operator \eqref{A_grvtn2}
which is dual to a fluctuation of the metric on $T^4$.  For this
operator $h=\tilde h=1$.  In this case, the summation over $k$ in
\eqref{AASigma} is the same as in (\ref{conicalG}). Therefore the
correlation function can be written as
\begin{align}
 G(w,\wb)&\equiv
 \bracket{\p X\pb X(w_1)\,\p X\pb X(w_2)}_\Sigma\notag\\
 &=-{1\over N}\sum_{n=1}^\infty
 {n N_n\over 16n^2\sin^2{w-\wb\over 2n}}
 \left[
 {1\over \sin^2{w\over 2}}+{1\over \sin^2{\wb\over 2}}
 - {2\sin{w-\wb\over 2}\over n\tan{w-\wb\over 2n}\sin{w\over 2}\sin{\wb\over 2}}
 \right].\label{dXdX_dXdX}
\end{align}
In Lorentzian signature we set
\begin{align}
 w=\phi-t,\qquad \wb=\phi+t \, .
\end{align}
The correlator $G(w,\wb)=G(t,\phi)$ then diverges at $w=k\pi/2$ or
$\wb=k\pi/2$ with $k\in \mathbb{Z}$.   This divergence is a
physical one, since on a finite cylinder a particle periodically
returns to the same spatial location.  Therefore, in order to make
the temporal behavior of the correlation function more
transparent, it is useful to remove this divergence.  So, let us
define the regularized correlator $\Gh(t,\phi)$ by dividing
$G(t,\phi)$ by the vacuum correlation function of the probe graviton operator:
\begin{align}
 \Gh(t,\phi)&\equiv -16\sin^2{w\over 2}\sin^2{\wb\over 2}\, G(t,\phi)\notag\\
 &={1\over N}\sum_{n=1}^\infty {n N_n\over (n\sin{t\over n})^2}
 \left[\sin^2{w\over 2}+\sin^2{\wb\over2}-{2\sin t\sin{w\over 2}\sin{\wb\over 2}\over n\tan{t\over n}}\right].
 \label{def_Ghat}
\end{align}

\begin{figure}[htb]
\begin{center}
 \begin{quote}
\begin{tabular}{c@{~~~~~~~~}c}
  \epsfxsize=7cm \epsfbox{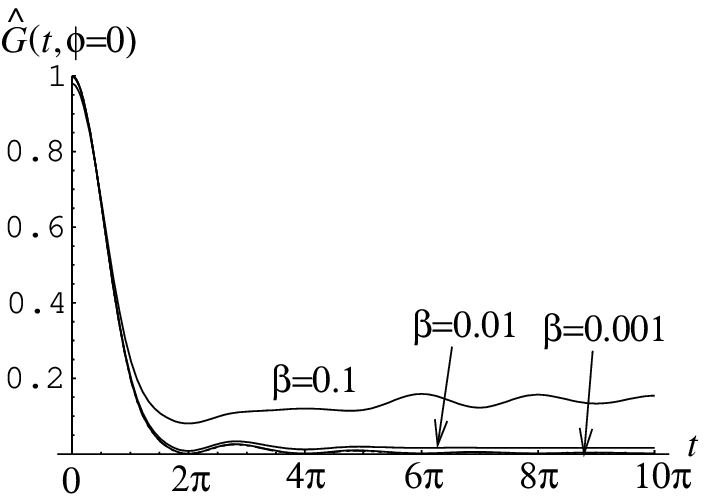}
 &
  \epsfxsize=7cm \epsfbox{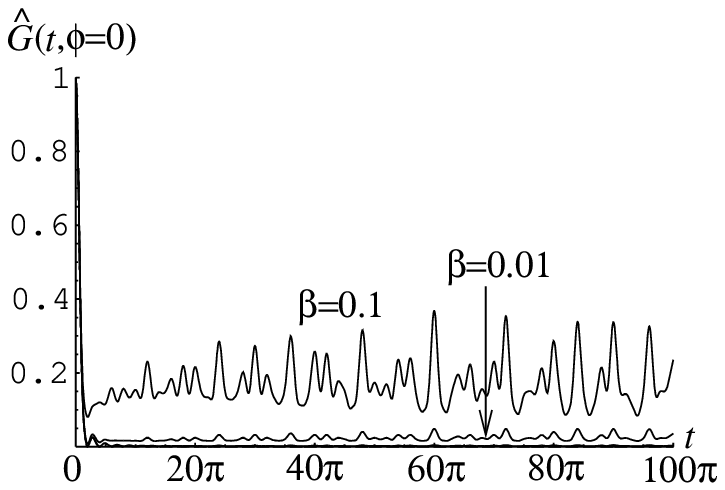}
     \\
  \epsfxsize=7cm \epsfbox{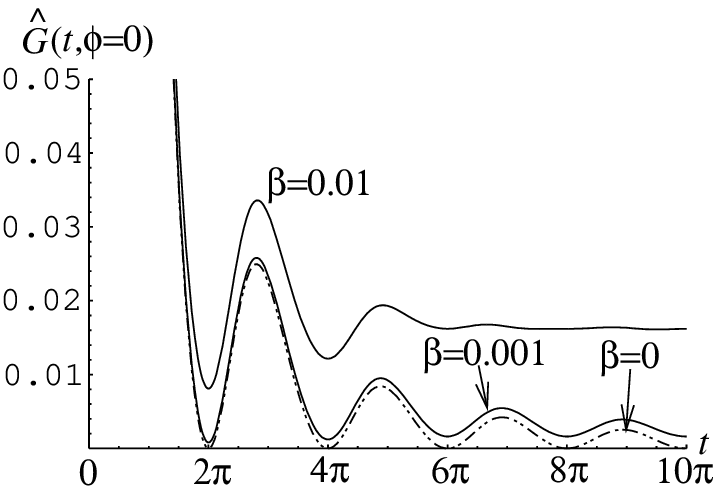}
 &
  \epsfxsize=7cm \epsfbox{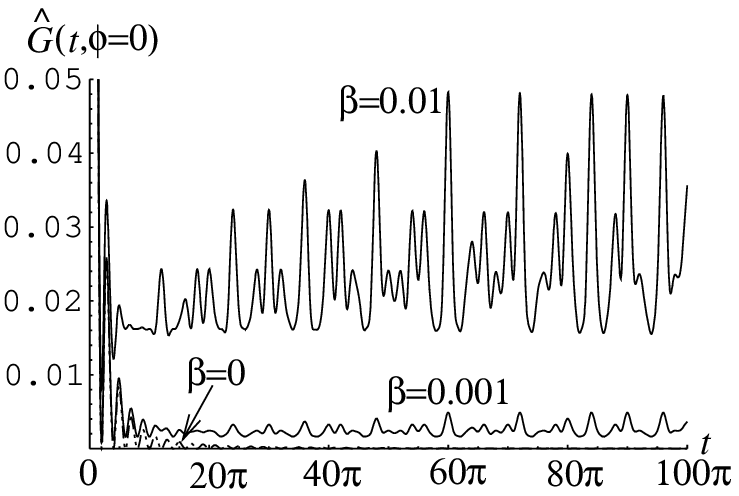}
\end{tabular}
 \caption{\sl Plot of the regularized correlation function
 $\Gh(t,\phi=0;\beta)$ as a function of $t$, for various values of
 $\beta$.  The two graphs on the left show short-time behavior; the two on the right  show
 long-time
 behavior.  As $\beta\to0$ (or equivalently, $N\to\infty$), $\Gh$
 approaches the correlation function (\ref{naiveG}) in the $M=0$ BTZ
 geometry, denoted in the graph by dashed lines. }
 \label{Ghat}
  \end{quote}
\end{center}
\end{figure}

Plugging in the representative distribution of constituent twists
for microstates with $J=0$  \eqref{typ_J=0}  into the regularized
correlator \eqref{def_Ghat} we obtain Fig.\ \ref{Ghat}.  As one can
see from this graph, the correlator decays rapidly at initial times
($t\lesssim\pi$), and  at later times exhibits a quasi-periodic
behavior.  Quasi-periodicity is not surprising; it is expected on
general grounds in a   system with a finite number of degrees of
freedom.   Furthermore, one sees that, in the $\beta\to 0$  (or
equivalently $N\to \infty$) limit, $\Gh$ approaches a certain limit
shape.  As we will discuss below, the limit shape in the
$N\to\infty$ limit turns out to be the correlation function
\eqref{naiveG} in the $M=0$ BTZ geometry.

\subsection{Effective geometry of microstates with $J=0$}

Now consider the correlation function \eqref{AASigma} of a general
bosonic non-twist operator:
\begin{align}
 \bracket{\CA(w_1) \CA(w_2)}_{\Sigma}
 ={1\over N}\sum_{n} n N_n \sum_{k=0}^{n-1}
 {C \over \left[2 n \sin\left({w-2\pi k\over 2n}\right)\right]^{2h}
 \left[2n \sin\left({\wb-2\pi k \over 2n}\right)\right]^{2\tilde h}}~.\label{ibvm8Jul05}
\end{align}
Let us study the relative size of the contributions  to this from
terms with different $n$.  The contributions come multiplied by $n
N_{n}$, which is $8n\over \sinh\beta n$ for the typical  microstates
with $J=0$ (Eq.\ \eqref{typ_J=0}).  Because of the suppression by
the $\sinh \beta n$,  the values of $n$ that make substantial
contributions to the correlation function \eqref{AASigma} are $n
\lesssim 1/\beta\sim \sqrt{N}$.
% Also, ${8n\over \sinh \beta n}$ is finite as $n\to 0$.
Thus there are $O(\sqrt{N})$ twists
that make a significant contribution.   Now observe that for any
$\gamma < 1/2$,  the number of twists with $n \lesssim N^\gamma$ is
parametrically smaller than $\sqrt{N}$. Indeed, the ratio vanishes
as $N \to \infty$.  In this sense we can say that in the $N \to
\infty$ limit, (\ref{ibvm8Jul05}) is dominated by twists scaling
as $n \sim \sqrt{N}$.

%On the other hand, apart from the $\sinh (\beta n)$ factor, the
%remaining part of the summand grows linearly in $n$ for large $n$.
%Combinining these two observations one finds that the summation gets
%its dominant contribution from $n \sim \sqrt{N}$.

%Thus there are $O(\sqrt{N})$ twists that make a significant
%contribution.   Now observe that for any $\gamma < 1/2$,  the number
%of twists that are $n \lesssim 1/\beta\sim \sqrt{N}$ is
%parametrically smaller than $\sqrt{N}$. Indeed, the ratio vanishes
%as $N \to \infty$.  In this sense we can say that in the $N \to
%\infty$ limit, (\ref{ibvm8Jul05}) is dominated by twists scaling
%as $n \sim N^{1/2}$.
% for which $n
%N_n\sim 1/\beta \sim\sqrt{N}$.
% In other words, in typical states,
%the typical length of the constituent twist operators is $n\sim
%\sqrt{N}$.  Hence, the behavior of \eqref{ibvm8Jul05} for large
%$N$ is dominated by the terms with $n\sim\sqrt{N}$.
Next, for any $n\ge 1$, when $t \ll n$ we can approximate the sum on $k$ as
\begin{align}
 \sum_{k=0}^{n-1}
 {1 \over \left[2 n \sin\left({w-2\pi k\over 2n}\right)\right]^{2h}
 \left[2n \sin\left({\wb-2\pi k \over 2n}\right)\right]^{2\tilde h}}
 \approx
 \sum_{k=-\infty}^{\infty}
 {1 \over (w-2\pi k)^{2h} (\wb-2\pi k)^{2\tilde h}}\qquad
 (t\ll n),
\notag% \label{n_infty_lim_OAAO}
\end{align}
where we assumed $h+\tilde h={\rm even}$.

Putting the above statements together, we arrive at the following
conclusion: for sufficiently early times
\begin{align}
 t\ll t_c=\CO(\sqrt{N}),
\end{align}
the correlation function \eqref{ibvm8Jul05} can be approximated by
\begin{align}
 \bracket{\CA(w_1) \CA(w_2)}_{\Sigma}
 &\approx {1\over N}\sum_{n} n N_n
 \sum_{k=-\infty}^{\infty}
 {C \over (w-2\pi k)^{2h} (\wb-2\pi k)^{2\tilde h}}\notag\\
 &=
 \sum_{k=-\infty}^{\infty}
 {C \over (w-2\pi k)^{2h} (\wb-2\pi k)^{2\tilde h}}.\label{eff_cor_J=0bos}
\end{align}
 This is
precisely the bulk correlation function in the naive geometry, or the
 $M=0$ BTZ black hole (compare with (\ref{naiveG}) for
$h=\tilde{h}=1$). Therefore, in the orbifold CFT approximation, the
emergent effective geometry of the D1-D5 system is the $M=0$ BTZ
black hole geometry. The description in terms of this effective
geometry is valid until $t\sim t_c$, which goes to infinity as
$N\to\infty$.  In the special case $h=\tilde{h}=1$, the summation
\eqref{eff_cor_J=0bos} yields (\ref{naiveG}).  In this case, we
indeed saw in Fig.\ \ref{Ghat} that the $\beta\to 0$ limit of the
correlation function is given by (\ref{naiveG}) (or
\eqref{eff_cor_J=0bos}).

Notice that in (\ref{eff_cor_J=0bos}) the sum  over the twists $n$
factors out.   Thus, for $t< t_c$ we are showing that the
correlation function is largely independent of the detailed
microscopic distribution of twists.  It is this universal response
that reproduces the physics of the $M=0$ BTZ black hole.
After $t\sim t_c$, the approximation \eqref{eff_cor_J=0bos} breaks
down, and the correlation function starts to show random-looking,
quasi-periodic behavior (see Fig.\ \ref{Ghat}). The form of the
correlation function in this regime will depend on the precise form
of the individual microstate, no matter how close it is to the
representative state \eqref{typ_J=0}.

The $\beta \rightarrow 0$ limit corresponding to the $M=0$ BTZ black
hole yields a correlator which decays to zero at large times as $1/t^2$.
By contrast, the microstate correlators exhibit quasi-periodic
fluctuations around a nonzero mean value. Numerical analysis indicates
that this mean value scales as ${1 \over \sqrt{N}}$ for $h=\tilde
h=1$. For an ordinary finite size, finite temperature system, one
expects the mean value to be of order $e^{-c S}$ where $S$ is the
entropy and $c$ is of order $1$.  This behavior arises because typical
interactions can explore the entire phase space of the system.  The fact
that we observe power law rather than exponential behavior is likely to
be a result of working in the free orbifold limit of the CFT and probing
the system with only non-twist operators.  Under these conditions the
full space of states does not come into play in determining a
correlation function.  For example, the non-twist operators cannot see
the full structure of the microstate, for instance the relative phases
between different twist components, and so might be expected to exhibit
larger correlations at late times. For this reason it would be very
instructive to repeat our analysis for twist operators, although this is
technically much more challenging.

A finite $N$ microstate correlator will exhibit exact periodicity in
time because only a finite number of frequencies appear in the
Fourier expansion.  The frequencies are $\omega_n = {n\over N},~ n=
1, 2, \ldots , N$.   Let $L(N)$ denote the least common multiple of
$(1,2, \ldots, N)$.   The correlator is then periodic with period
$\Delta t = 2\pi N L(N)$.    The large $N$ behavior of $L(N)$ is
$L(N) \sim e^N$, and therefore
\begin{align}
\Delta t \sim N e^N~.
\end{align}
Our correlators have been computed in the canonical ensemble in
which the summation over $n$ extends past $N$ up to infinity, and so
we will not see this exact periodicity.  On the other hand, due to
the exponential suppression of the distribution function $N_n$ the
deviation from exact periodicity is tiny for large $N$. As was
argued above, and as can be confirmed numerically,  one finds that
for large $N$ the large time behavior of the correlator is
unaffected if we truncate the sum over $n$ at $n_{\rm max} = c
\sqrt{N}$, for $c$ of order unity. Taking this into account, we see
that our correlators will exhibit approximate periodicity with
period
\begin{align}
\Delta t \sim e^{c \sqrt{N}} = e^{\tilde{c} S} ~,
\end{align}
where $S=2\pi \sqrt{2}\sqrt{N}$ is the entropy.  This timescale is
the so-called Poincar\'{e} recurrence time, over which generic
finite size thermal systems are expected to exhibit approximate
periodicity.

\bigskip\bigskip
\noindent{\bf Fermionic probes: } In  the above we restricted
ourselves to  bosonic probes, but we obtain the same effective
geometry even if we probe the microstate with operators that contain
fermions.  For example, let $\CA=\Psi^{s'}\Psit^{\st'}$ as defined
in \eqref{bosonize}. From \eqref{ferm_corr_bldblk}, we obtain
\begin{align}
 \bracket{\CA_A\CA_B}_{\sigma_n^{s\st}}
 =
 \bracket{ [\Psi_A^{s'}\Psit_A^{\st'}(w_1)]^\dagger  \,
 \Psi_B^{s'}\Psit_B^{\st'}(w_2)}_{\sigma_n^{s\st}}
 &=
 {e^{iss'w/2n-i\st\,\st'\wb/2n}\over (2n\sin{w\over 2n})
 (2n\sin{\wb\over 2n})}~.\label{jdqt10Jun05}
\end{align}
If we sum over copies,
\begin{align}
 \sum_{A=1}^n \bracket{\CA_A\CA_1}_{\sigma_n^{s\st}}
 &=
 \sum_{k=0}^{n-1}
 {e^{iss'(w+2\pi k)/2n-i\st\,\st'(\wb+2\pi k)/2n}
 \over (2n\sin{w+2\pi k\over 2n})(2n\sin{\wb+2\pi k\over 2n})}\notag\\
 &\approx
 \sum_{k=-\infty}^{\infty}
 {1 \over (w+2\pi k)(\wb+2\pi k)}\qquad (t\ll n).
 \label{iabi10Jun05}
\end{align}
Other twist operators
$\sigma_n^{\alphat\betat},\tau_n^{s\alphat},\tau_n^{\alphat\st}$
give different correlation functions, but they are all identical
to \eqref{iabi10Jun05} for $t\ll n$:
\begin{align}
 \sum_{A=1}^n \bracket{\CA_A\CA_1}_{\sigma_n^\muh}
 \approx
 \sum_{k=-\infty}^{\infty}
 {1 \over (w+2\pi k)(\wb+2\pi k)}
 \qquad (t\ll n)
 ,\label{jkeg10Jun05}
\end{align}
where $\sigma_n^\muh$ can be any of the twist operators
$\sigma_n^{s\st}, \sigma_n^{\alphat\betat}, \tau_n^{s\alphat},
\tau_n^{\alphat\st}$.  Plugging this result into \eqref{SAAS2}, we
conclude that
\begin{align}
 \bracket{\CA(w_1) \CA(w_2)}_{\Sigma}
 &=
  \bracket{ [\Psi_A^{s'}\Psit_A^{\st'}(w_1)]^\dagger  \,
 \Psi_B^{s'}\Psit_B^{\st'}(w_2)}_\Sigma
 \approx
 \sum_{k=-\infty}^{\infty}
 {1 \over (w-2\pi k) (\wb-2\pi k )}
 \qquad (t\ll t_c).
\end{align}
This is again the correlation  function in the $M=0$ BTZ black hole
geometry. Therefore, we conclude that the  effective geometry of the
D1-D5 system in the orbifold CFT approximation is the $M=0$ BTZ
black hole geometry for any non-twist probe operators, bosonic or
fermionic.  The description by this effective geometry breaks down
at $t=t_c=\CO(\sqrt{N})$.

\bigskip\bigskip
\noindent{\bf Gravitational origin of the effective geometry: } We
can also argue that the effective geometry for  the ensemble with
$J=0$ should be the  $M=0$ BTZ black hole by using the Lunin--Mathur
metric \eqref{LMmetric1}.    Assume that the profile $F_i(v)$ is a
random superposition of small-amplitude, high-frequency oscillations
that is much smaller than the asymptotic AdS radius:
\begin{align}
 |\Fv(v)|\ll \ell\sim N^{1/4}.\label{kjcj8Jul05}
\end{align}
Then, for $r\sim\ell\gg |\Fv(v)|$,
\begin{align}
 f_5 &={Q_5\over L}\int_0^L dv {1\over |\xv-\Fv(v)|^2} \approx{Q_5\over r^2},\\
 f_1 &={Q_5\over L}\int_0^L dv {|\dot\Fv(v)|^2\over |\xv-\Fv(v)|^2}
 \approx{1\over r^2}{Q_5\over L}\int_0^L dv\, {|\dot\Fv(v)|^2} ={Q_1\over r^2},\\
 A_i &={Q_5\over L}\int_0^L dv {\dot F_i(v)\over |\xv-\Fv(v)|^2}
 \approx0,
\end{align}
where in the second line we used \eqref{Q12}, and in the third line
$A_i(x)$ vanishes because $F_i(v)$ is random.  So the metric
\eqref{LMmetric1} is
\begin{align}
 ds^2&=-{r^2\over \ell^2}dt^2+{r^2\over \ell^2}dy^2
 +{\ell^2\over r^2}(dr^2+r^2d\Omega_3^2)+\sqrt{Q_1\over Q_5}\,ds_{T^4}^2
\end{align}
with $\ell=(Q_1 Q_5)^{1/4}$.  This is indeed the direct product of
the $M=0$ BTZ black hole \eqref{naive}, and $S^3\times T^4$, if one
sets $y\to R\phi$, $r\to\ell r/R$.  One can check that the condition
\eqref{kjcj8Jul05} is satisfied from the microscopic theory, as
follows. If the typical frequency and amplitude are $\omega$ and
$a$, respectively, then $|\Fv|\sim a$, $|\dot\Fv|\sim a\omega$.  We
can relate $\omega,a$ with the microscopic quantities $n,N_n$ as
$\omega\sim n$, $a\sim N_n^{1/2}$.  Recall that the typical twist is
$n\sim N^{1/2}\sim 1/\beta$.  For $n\sim1/\beta$, $N_n$ is
$N_n={8\over \sinh(\beta n)}=\CO(1)$ from \eqref{typ_J=0}.
Therefore, $\omega\sim N^{1/2}, a\sim N^0$.  This indeed satisfies
\eqref{kjcj8Jul05}.

\subsection{Effective geometry of microstates with $J\neq 0$}

As we saw in subsection \ref{subsec:typ_J=/=0}, the ensemble with $J\neq
0$ becomes in the large $N$ limit a ``direct product'' of the
Bose--Einstein condensate $(\sigma_{1}^{s\st})^{|J|}$ with $s=\st=\mp$,
and an ensemble with level $\Nt=N-|J|$ and no angular momentum.
Therefore, from the general formula \eqref{SAAS2},  one sees that
the correlation function for this ensemble is a sum of the
correlation function in the Bose--Einstein condensate background and
the one for the ensemble with level $\Nt=N-|J|$ and no angular
momentum.

Specifically, consider a bosonic non-twist operator $\CA$.
Plugging the typical distribution \eqref{typ_J=/=0} into the
formula \eqref{AASigma},
\begin{align}
 &\bracket{\CA(w_1) \CA(w_2)}_{\Sigma}\notag\\
 &=
 {|J|\over N}{C \over \left(2\sin{w\over 2}\right)^{2h}
 \left(2 \sin{\wb\over 2}\right)^{2\tilde h}}
 +{1\over N}\sum_{n} n \tilde N_n \sum_{k=0}^{n-1}
 {C \over \left[2 n \sin\left({w-2\pi k\over 2n}\right)\right]^{2h}
 \left[2n \sin\left({\wb-2\pi k \over 2n}\right)\right]^{2\tilde h}}
 \notag\\
 &\approx
 {|J|\over N}{C \over \left(2\sin{w\over 2}\right)^{2h}
 \left(2 \sin{\wb\over 2}\right)^{2\tilde h}}
 +\left(1-{|J|\over N}\right)
 \sum_{k=-\infty}^{\infty}
 {C \over ({w-2\pi k})^{2h}(\wb-2\pi k)^{2\tilde h}}
 \qquad (\boldsymbol{t\ll t_c})
 \notag\\
 &=
 {|J|\over N} \bracket{\CA\CA}_{\rm BEC}
 +
 \left(1-{|J|\over N}\right)\bracket{\CA\CA}_{\text{$M=0$ BTZ}}~.\label{gmwl9Jul05}
\end{align}
The critical time $t_c$ is now given by
\begin{align}
 t_c&=\CO(\sqrt{N-|J|}).\label{tc_J=/=0}
\end{align}
The first term in \eqref{gmwl9Jul05}, which arises from the
Bose--Einstein condensate (BEC), is proportional to the correlation
function of $\CA$ computed in global $\ads{3}$.
This is happening because the condensate is $(\sigma_1^{s\st})^{|J|}$,
$s=\st=\mp$, and the 3-dimensional part of the microstate geometry
associated with this operator by itself is simply global $\ads{3}$ with
a scale $\ell \sim |J|^{1/4}$, as described by
\cite{Balasubramanian:2000rt,Maldacena:2000dr, Lunin:2002iz}.  Actually
the total 10-dimensional geometry is more complicated because of the
nontrivial Wilson line coming from the internal $S^3$, but the bosonic
operator $\CA$ does not sense this extra structure.  On the other hand,
fermionic $\CA$ does see this structure, as we will see below.
%\footnote{There is
%also a Wilson line that is not sensed by the bosonic operator $\CA$.}
%

Hence the ``effective geometry'' for $t<t_c$ appears to be a weighted
average of global $\ads{3}$ (with a nontrivial Wilson line) and the
$M=0$ BTZ\@. The linear summation in \eqref{gmwl9Jul05} is appearing
because in the orbifold CFT the simple class of non-twist probes has
correlation functions that are simply linear summations of the responses
in the individual constituent twist states (\ref{gmwl9Jul05}).  Of
course as $|J| \to N$, the typical microstate operator found in
(\ref{hlro26May05}) becomes precisely the operator corresponding to
global $\ads{3}$ (with a Wilson line) in \cite{Balasubramanian:2000rt,
Maldacena:2000dr, Lunin:2002iz}. Thus the response (\ref{gmwl9Jul05}) is
simply a weighted sum of the expected responses in the $J=0$ and $|J| =
N$ limits.
%It is worth recalling that (\ref{hlro26May05}) is simply the
%dominant piece of the typical microstate operator.  Individual
%microstates will differ from this operator both in the ``condensate''
%part and in the remaining part that carries the entropy.  Fluctuations
%in the former when $|J| < N$ will smear the clean identification of the
%first term in (\ref{gmwl9Jul05}) as the $\ads{3}$ correlator.

Correlation functions involving fermionic operators can be evaluated
similarly.  For example, let us take $\CA=\Psi^{s'}\Psit^{\st'}$ as
before.  From \eqref{AASigma}, \eqref{jdqt10Jun05}, and
\eqref{jkeg10Jun05}, we obtain
\begin{align}
 \bracket{\CA_A\CA_B}_{\Sigma}
 &\approx
 {|J|\over N}
 {e^{is(s'w-\st'\wb)/2}\over (2\sin{w\over 2})(2\sin{\wb\over 2})}
 +\left(1-{|J|\over N}\right)
 \sum_{k=-\infty}^{\infty}
 {1 \over ({w-2\pi k})(\wb-2\pi k)}
 \qquad (t\ll t_c)
 \notag\\
 &=
 {|J|\over N} \bracket{\CA\CA}_{\text{BEC}}
 +
 \left(1-{|J|\over N}\right)\bracket{\CA\CA}_{\text{$M=0$ BTZ}} ,\label{gmyj9Jul05}
\end{align}
where $s={\rm sign}(J)$.  Again the ``effective geometry'' appears to be
a weighted average.

The Bose--Einstein condensate part $\bracket{\CA\CA}_{\text{BEC}}$ of the
bosonic correlator \eqref{gmwl9Jul05} did not care whether the
condensate is made of $\sigma_1^{s\st}$ with $s=\st=-1$ or $s=\st=+1$,
whereas the fermionic one \eqref{gmyj9Jul05} does depend on what the
condensate is made of through its dependence on $s={\rm sign}(J)$.  This
reflects the fact that the 3-dimensional geometry corresponding to
$(\sigma_1^{s\st})^N$ with $s=\st=-1$ and the one with $s=\st=+1$ are
both global AdS$_3$ but differ in the nontrivial Wilson line in the
internal $S^3$ \cite{Balasubramanian:2000rt, Maldacena:2000dr,
Lunin:2002iz}. Bosonic probes are not charged under the relevant $U(1)$,
and thus its correlator is independent of what the condensate is made
of. On the other hand, fermionic probes are charged under the $U(1)$,
and its correlator depends on what the condensate is made of.

%\bigskip
%
Do the above results mean that the emergent geometry is a superposition
of two classical geometries?  Below we will use the Lunin-Mathur
solution \eqref{LMmetric1} to argue that this should not be the
case and that the emergent geometry should be a singular zero-horizon
limit of the black ring \cite{Elvang:2004rt}.

\bigskip\bigskip

\noindent{\bf The effective geometry should be a black ring: }  Assuming $J>0$ and $J=\CO(N)$, the typical state of the ensemble with $J\neq 0$ is given by \eqref{hlro26May05}.  In the  language
of the FP system, $(\alpha_{-1}^+{}^\dagger)^J$ corresponds to an
F1 worldvolume that makes a circle with radius
$\sim\sqrt{J}=\CO(N^{1/2})$ in the 1-2 plane.  The remaining part
$\prod_{n=1}^\infty
\left[\prod_{i}(\alpha_{-n}^{i})^{N_{ni}}(\psi_{-n}^{i})^{N'_{ni}}\right]$
adds fluctuations around this circular profile.  By an argument
similar to the one given at the end of the last subsection, the
typical frequency and amplitude of the fluctuations are estimated
to be $n\sim\sqrt{N-J}=\CO(N^{1/2})$ and $N_n^{1/2}=\CO(N^0)$,
respectively.

This motivates the following profile function $\Fv(v)$ of  the
D1-D5 metric \eqref{LMmetric1}.  Namely, we assume that the profile
$\Fv(v)$ is a circle $\Fv^{(0)}$ with random, small-amplitude,
high-frequency fluctuations $\delta \Fv$ around it:
\begin{align}
 \Fv&=\Fv^{(0)}+\delta\Fv,\qquad
 \begin{cases}
 F_1^{(0)}+i F_2^{(0)}=a e^{i \omega v},\\
  F_3^{(0)}=F_4^{(0)}=0,
 \end{cases}\qquad
 \omega={2\pi\over L}={R\over Q_5}.\label{iumd28Jun05}
\end{align}
From the above analysis, the amplitude of the fluctuation $\delta \Fv$
is much smaller than the size of the circle or the AdS radius:
\begin{align}
 |\delta \Fv|=\CO(N^0)\ll |\Fv^{(0)}|=a=\CO(N^{1/2}),
 \qquad
 |\delta \Fv|=\CO(N^0) \ll\ell=\CO(N^{1/4}).
\end{align}
On the other hand, the derivatives of $\Fv^{(0)}$ and $\delta\Fv$ are of
the same order of magnitude:
\begin{align}
 |\delta\dot \Fv|\sim n N_n^{1/2}=\CO(N^{1/2}), \qquad |\dot\Fv^{(0)}|=a\omega=\CO(N^{1/2}) .
\end{align}
Using these relations, the harmonic functions in \eqref{LMmetric1} are
approximated for large $N$ as follows:\footnote{This metric was studied
in \cite{Lunin:2002bj} using a different ansatz of the profile function
$\Fv(v)$.  Recent analysis of this metric from the bubbling AdS
viewpoint of \cite{Lin:2004nb} can be found in \cite{Boni:2005sf}.}
\begin{align}
 f_5&\approx {Q_5\over L}\int_0^L dv{1\over |\xv-\Fv^{(0)}|^2} ={Q_5\over\Sigma},\notag\\
 f_1&\approx {Q_5\over L}
 (a^2\omega^2+|\delta\dot \Fv|^2) \int_0^L dv
   {1\over |\xv-\Fv^{(0)}|^2}={Q_1\over\Sigma},\label{harm_J=/=0}\\
 A_1+i A_2&\approx {Q_5\over L}\int_0^L dv{ia\omega e^{i\omega v}\over
 |\xv-\Fv^{(0)}|^2},
 \qquad {\rm therefore}\quad
 A_\psi={2a^2Q_5\omega s^2\over \Sigma(\Sigma+s^2+w^2+a^2)},
 \notag
\end{align}
where
\begin{align}
 x_1+i x_2=s e^{i\psi},\quad x_3+i x_4=w e^{i\phi},\quad
 \Sigma=\sqrt{[(s+a)^2+w^2][(s-a)^2+w^2]}.
 \label{iaua19Jul05}
\end{align}
In the second line of \eqref{harm_J=/=0}, the cross term
$\Fv^{(0)}\cdot \delta\Fv$ was dropped because $\delta\Fv$ is
fluctuating randomly.
Also in the second line, because $|\delta\dot \Fv|^2$ is
fluctuating with length scale much smaller than $a$, we can
replace it with it average and take it out of the integral (so,
$|\delta\dot \Fv|^2$ there really means the average).  We also
used the relation \eqref{Q12}:
\begin{align}
 Q_1&={Q_5\over L}\int_0^L dv |\dot\Fv|^2\approx(a^2\omega^2+|\delta\dot\Fv|^2)Q_5.
\end{align}
In the third line of \eqref{harm_J=/=0}, the term containing $\delta\dot
\Fv$ was dropped because it is fluctuating randomly.
It is convenient to go to the $(x,y,\psi,\phi)$ coordinate system
\cite{Elvang:2004rt} with $R=a$, defined by
\begin{align}
 s&={\sqrt{y^2-1}\over x-y}R,\qquad
 w={\sqrt{1-x^2}\over x-y}R.
\end{align}
In this coordinate system, $A_i$, $B_i$, $\Sigma$ can be written as
\begin{align}
 A_\psi={Q_5\omega\over 2}(-1-y),\qquad
 B_\phi={Q_5\omega\over 2}(1+x),\qquad
 \label{iojs28Jun05}
 \Sigma={2R^2\over x-y}.
\end{align}
Plugging these into \eqref{LMmetric1}, one obtains the metric
\begin{align}
 ds^2&={\Sigma\over \ell^2}
 \left[
 -(dt+{Q_5\omega\over 2}(-1-y)d\psi)^2
 +(dy+{Q_5\omega\over 2}(1+x)d\phi)^2
 \right]
 +{\ell^2\over\Sigma}ds_4^2+\sqrt{Q_1\over Q_5}ds_{T^4}^2,\label{sing_BR}
\end{align}
where $\ell\equiv (Q_1Q_5)^{1/4}$.  This is the metric  of the
supersymmetric black ring \cite{Elvang:2004rt,Bena:2004de} with
charges $(Q_1,Q_2,Q_3)=(Q_1,Q_5,0)$, dipole charges
$(q_1,q_2,q_3)=(0,0, Q_5\omega)$, and radius $R=a$.  For these
charges, the horizon area and thus the Bekenstein--Hawking entropy
vanish.  The angular momentum of this singular black ring
satisfies
\begin{align}
 J_\psi&=R^2q_3\le {Q_1 Q_2\over q_3}\equiv J_{\psi,\rm max}.\label{D1D5kk_reg}
\end{align}
If this inequality is saturated, the singular black ring becomes the
regular D1-D5$\to$kk geometry.  However, in the present case,
\begin{align}
 J_\psi=a^2 Q_5 \omega,\qquad
 J_{\psi,\rm max}=
 {Q_1Q_5\over Q_5\omega}
 =
 a^2Q_5\omega \left(1+{|\delta\dot\Fv|^2\over a^2\omega^2}\right).
\end{align}
So, the equality in \eqref{D1D5kk_reg} does not hold and the geometry
\eqref{sing_BR} describes a singular, zero-horizon limit of the black
ring.

The above argument suggests that the effective geometry for the ensemble
with $J\neq 0$ is the singular, zero-horizon limit of the black ring
\eqref{sing_BR}.\footnote{This is reminiscent of the proposal by
\cite{Bena:2004tk} that the CFT microstate of the black ring with
non-vanishing horizon is made of two parts, where the first part is made
of small effective strings of identical length, while the second part is
made of a single long string and responsible for the whole entropy.}
The description by this effective
geometry should be valid up to the critical time $t_c$ \eqref{tc_J=/=0},
which goes to infinity as $N\to \infty$.
In order to prove the above statement, one should compute the
bulk-boundary propagator in the singular black ring geometry
\eqref{sing_BR} and show that it leads to the boundary CFT
correlation function \eqref{gmwl9Jul05}, \eqref{gmyj9Jul05}.

\section{Discussion}
\label{sec:disc}

The puzzles regarding the black hole information paradox are all
traceable to the fact that we don't have an adequate understanding
of the relation between geometry and entropy.  In the boundary CFT
description of black holes we can choose to work either with
individual microstates or with an ensemble, and we understand that
entropy arises from the coarse-graining associated with defining
the ensemble.  In practice the ensemble usually yields results to
the accuracy we desire, and the existence of the underlying
microstate description tells us that there is no possibility of
information loss at a fundamental level.

We lack a similar understanding in the bulk.  If the black hole is
to be thought of as an ensemble, we need to specify precisely the
elements of the ensemble.  One logical possibility is that the
bulk description is intrinsically coarse-grained, and that
microstates can only be found in the boundary CFT\@.  An alternative
picture, advocated by Mathur, is that bulk microstates are to be
described as new horizon-free geometries differing from the black
hole at the horizon scale.  Some evidence for the latter has
accumulated, but the question remains open.

Here, we have studied some of these issues in the simple context of
the D1-D5 CFT at the free orbifold point.  On the one hand,  a large
class of microstate geometries are known, and on the other hand
there is an effective ``black-hole'' geometry describing their
``average''.   We essentially tried to make this last sentence
precise by comparing CFT correlation functions computed in typical
microstates to bulk correlation functions computed in the
``black-hole'' geometry.   The agreement we found, as well as its
breakdown at late times, provides evidence for the picture of black
holes as the effective description of more fundamental underlying
structures.   Although the ``black-hole'' in this case has vanishing
horizon size, it does display some of the hallmarks of real black
holes, such as the decay of late time correlators.

If black holes in general represent effective coarse-grained descriptions
of underlying microstate geometries, it naturally explains why one cannot
see quasi-periodicity and Poincar\'e recurrence by summing over the
$SL(2,\mathbb Z)$ family of BTZ black holes as was pursued in
\cite{Barbon:2003aq,Kleban:2004rx}.  This is analogous to the fact that,
after replacing a gas of molecules by its effective coarse-grained
description, {\it i.e.}\ a dissipative continuum, one does not expect to
be able to see quasi-periodicity or Poincar\'e recurrence in the
correlation function describing a particle  scattered in the
gas.

It would be interesting to try to repeat our calculations in the context
of the D1-D5 system on K3 rather than $T^4$.  In the K3 case it has been
found that higher derivative terms in the supergravity Lagrangian lead
to a nonzero size horizon whose Bekenstein--Hawking--Wald entropy agrees
with that of the CFT \cite{Dabholkar:2004yr,Dabholkar:2004dq}.
Furthermore, one can still write down a large class of microscopic
geometries which contribute to the entropy \cite{Giusto:2004xm}.  The
complication is that the sigma model is no longer free, and so the
computation of CFT correlators is not as straightforward. But the goal
would be to show how the nonzero horizon size manifests itself in CFT
correlators.  Alternatively, perhaps a horizon could be found even in
the $T^4$ case once interactions are included.

Another useful endeavor would be to compare bulk correlators computed in
the known microstate geometries of the D1-D5 system to the microscopic
CFT correlators we have computed here.  This easily can be done for the
simplest class of states, namely those corresponding to the twist
operator $\sigma=[\sigma_n^{s\st}]^{N/n}$, $s=\st=-1$.  In this case the bulk
geometries are simply the conical defects (\ref{condef}), and we saw
that this gives precise agreement between bulk and boundary correlators.
But for more general states the bulk geometry is no longer just an
orbifold, and the bulk correlators will be much more complicated.  On
the other hand, the CFT correlators continue to be expressed as a sum of
simple contributions.  This suggests that either the bulk geometries can
also somehow be thought of as being built up out of simple geometries,
or alternatively that working at the free orbifold point of the CFT is
simply inadequate.

In this paper we studied correlation functions of non-twist
operators, but it would be very interesting to consider twist
operators.  This would allow much greater sensitivity to the
microstate structure.  Non-twist operators see the states as built
up out of decoupled components corresponding to the given cycles,
and this led to the correlators taking the form of a sum over
relatively simple contributions from each component.   This will
no longer be the case when twist operators are used to probe the
state, and the results are expected to be much more complicated.
This extra information could potentially be used to map out the
bulk geometry in much greater detail.

\section*{Acknowledgments}

We would like to thank Jan de Boer, Hiroshi Fujisaki, Norihiro Iizuka,
Vishnu Jejjala, Oleg Lunin, Joan Simon, Sanefumi Moriyama, and Hirosi
Ooguri for valuable discussions.
We would also like to thank the organizers of the workshop on
Quantum Theory of Black Holes at the Ohio State University, where
this work was initiated, the Workshop on Gravitational Aspects of
String Theory at the Fields Institute, and Strings 2005, for
stimulating environments.
M.S. would like to thank Norihiro Iizuka for collaboration in
\cite{Iizuka:2005uv} and helpful discussions, and the Theoretical High
Energy Physics group at the University of Pennsylvania for hospitality.
P.K. was supported in part by NSF grant PHY-0099590.  M.S. was supported
in part by Department of Energy grant DE-FG03-92ER40701 and a Sherman
Fairchild Foundation postdoctoral fellowship.  V.B. was supported in
part by the DOE under grant DE-FG02-95ER40893, by the NSF under grant
PHY-0331728 and by an NSF Focused Research Grant DMS0139799.

\appendix

\section{D1-D5 CFT}
\label{app:cft}

In this appendix we present a complete review of the relevant aspects of
the D1-D5 CFT, in particular chiral primary fields and the corresponding
R(amond) ground states related by spectral flow.  We will compute
correlation functions of non-twist operators in the R ground states,
which are related via AdS/CFT to supergravity amplitudes in AdS$_3
\times S^3$.
References on $S_N$ orbifold CFTs and methods for
computing correlation functions in them include \cite{Hamidi:1986vh,
Dixon:1986qv, Bantay:1997ek, Arutyunov:1997gt, Evslin:1999qb,
Jevicki:1998bm, Mihailescu:1999cj, Lunin:2000yv, Argurio:2000tb,
Lunin:2001pw, deBoer:2001nw}. 
Below we will closely follow the argument of
\cite{David:2002wn,Wadia:2000wy,Lunin:2001pw} and the notation of
\cite{Lunin:2001pw}.  For a more detailed explanation of the covering
space method and the NS sector chiral primaries, see \cite{Lunin:2000yv,
Lunin:2001pw}.

The main results from this appendix that are used in the main text of
this the paper are the bosonic two-point function (\ref{corrAA_w}) and
the fermionic two-point functions
(\ref{fermia})--(\ref{ferm_corr_bldblk2}).  The two-point function for
the general state \eqref{gen_twist} can be computed using \eqref{SAAS}
and \eqref{4pt_func}.  We will derive these using orbifold CFT
machinery, but the final results for the two-point function are simple
and intuitive, and can be obtained more simply by just taking into
account the fact that the effective length of the CFT cylinder undergoes
a rescaling.  However, the detailed machinery described below is
necessary for more computation of more general correlation functions,
particularly those involve twist operators as probes.

\subsection{D1-D5 system}
\label{app:cft:d1d5sys}

Consider type IIB string theory on $\mathbb  R_t\times \mathbb
R^{4} \times S^1 \times T^4$ with $N_1$ D1-branes and $N_5$
D5-branes. The D1-branes are wound on $S^1$ and smeared over
$T^4$, and the D5-branes are wrapped on $S^1\times T^4$ .  We
denote by $x^0$ the time direction $\mathbb R_t$; by $x^i$
$(i=1,2,3,4)$ the $\mathbb R^{4}$ directions; by $x^5$ the $S^1$
direction; and by $x^a$ ($a=6,7,8,9$) the $T^4$ directions (see
Table \ref{D1D5config}).
\begin{table}[htb]
 \begin{center}
 \begin{tabular}{|c|cccccccccc|}
  \hline
     & 0 & 1 & 2 & 3 & 4 & 5 & 6 & 7 & 8 & 9\\
  \hline
  D1 & $\bigcirc$ & $\cdot$ & $\cdot$ &  $\cdot$ & $\cdot$ & $\bigcirc$ & $\sim$ & $\sim$ & $\sim$ & $\sim$\\
  D5 & $\bigcirc$ & $\cdot$ & $\cdot$ & $\cdot$ & $\cdot$ & $\bigcirc$ & $\bigcirc$ & $\bigcirc$ & $\bigcirc$ & $\bigcirc$\\
    \hline
 \end{tabular}
 \end{center}
 \vspace*{-3ex}
 \caption{\sl Configuration of D-branes}
 \label{D1D5config}
\end{table}
The low energy world-volume dynamics of the D1-D5 system is described by
a $(1+1)$-dimensional $\CN=(4,4)$ SCFT in the RR sector, where the two
dimensions come from the $x^{0,5}$ directions
\cite{Strominger:1996sh,deBoer:1998ip,Seiberg:1999xz,Larsen:1999uk}.
This theory has $SO(4)_E\cong SU(2)_R\times \widetilde{SU(2)}_R$
$R$-symmetry, which originates from the rotational symmetry in the
transverse directions $x^{i}$, $i=1,2,3,4$.  On the other hand, the
rotation in the longitudinal directions $x^a$, $a=6,7,8,9$ leads to
$SO(4)_I\cong SU(2)_{I}\times \widetilde{SU(2)}_I$ symmetry.  Actually,
$T^4$ breaks the latter symmetry, but it can still be used for
classifying the states in the theory.

The CFT is a sigma model whose target space is the symmetric product
$\CM_0=(T^4)^N/S_N$, where $S_N$ is the permutation group of order $N$.
We put
\begin{align}
 N=N_1 N_5.
\end{align}
More precisely, the target space is not the symmetric product
$\CM_0$ but a deformation of it; the sigma model has marginal
deformations, which one has to turn on in order for the CFT to
precisely correspond to the supergravity side.  $\CM_0$ is a
special point in the  moduli space of the CFT called the orbifold
point, where the CFT becomes free.  This situation is very similar
to the situation of AdS$_5$/SYM$_4$ duality, where SYM becomes
free at a special point ($g_{\rm YM}=0$) in the moduli space, but
in order for SYM to precisely correspond to the supergravity side
one has to turn on the coupling $g_{\rm YM}$.  The orbifold point
is the analogue of the free SYM\@.  In the following, we will
consider the orbifold point of the D1-D5 CFT.
\subsection{Orbifold CFT}\label{app:cft:orb_cft}

The $\CN=(4,4)$ SCFT at the orbifold point ${\CM_0}=(T^4)^N/S_N$ is
described by the free Lagrangian
\begin{align}
 S&={1\over 2\pi}\int d^2\sigma [\partial x_A^a\pb x^a_A
 +\psi_A^a(z)\pb\psi_A^a(z)+\psit_A^a(\zb)\p\psit^a_A(\zb)],\label{S_orb}
\end{align}
where $a=6,7,8,9$ labels the $T^4$ directions  and $A=1,\cdots,N$
labels the $N$ copies of $T^4$.  Summation over $a$ and $A$ is
implied. Without the orbifolding, this theory would be simply a
direct sum of $N$ free CFTs each with $c=6$.

As we explained in the last subsection, this theory has $SO(4)_E\cong
SU(2)_R\times \widetilde{SU(2)}_R$ $R$-symmetry and $SO(4)_I\cong
SU(2)_I\times \widetilde{SU(2)}_I$ non-$R$-symmetry.  The transformation
property of the fields under these symmetry groups is as
follows:\footnote{The surviving supersymmetry is in the representation
$(+\half;{\bf 2,1;2,1})$ and $(-\half;{\bf 1,2;2,1})$ under
$SO(1,1)_{05}\times [SU(2)_R\times\widetilde{SU(2)}_R] \times
[SU(2)_I\times\widetilde{SU(2)}_I]$.  This implies the transformation
property \eqref{chg_tbl} of the hypermultiplet superpartners of the
boson $x^a$ (see e.g.\ \cite{Maldacena:1999bp}).}
\begin{align}
 \renewcommand{\arraystretch}{1.3}
\begin{array}{|c|c|c|c|}
 \hline
 \text{field}
 & SU(2)_R\times \widetilde{SU(2)}_R \rule{0pt}{3ex}
 & SU(2)_I\times \widetilde{SU(2)}_I \rule{0pt}{3ex}\\
 \hline
 \hline
 x^a       & {\bf (1,1)} & {\bf (2,2)}\\ \hline
 \psi^a    & {\bf (2,1)} & {\bf (1,2)}\\ \hline
 \psit^a   & {\bf (1,2)} & {\bf (1,2)}\\ \hline
% G^\alpha       & {\bf (2,1)} & {\bf (1,2)}\\ \hline
% \tilde G^\alphat& {\bf (1,2)} & {\bf (2,1)}\\ \hline
\end{array}
 \label{chg_tbl}
\end{align}
Following \cite{Lunin:2001pw}, we bosonize the fermions as
\begin{align}
 \Psi^+_A(z)&\equiv{1\over \sqrt{2}}(\psi^1_A+i\psi^2_A)=e^{ i\phi^5_A(z)},\qquad
 \Psi^-_A(z) \equiv{1\over \sqrt{2}}(\psi^3_A+i\psi^4_A)=e^{ i\phi^6_A(z)},\label{bosonize}
\end{align}
where left-moving bosons are normalized as
$\phi^i(z_1)\phi^j(z_2)\sim -\delta^{ij}\log(z_1-z_2)$.
Similarly, the right moving fermions $\Psit^\pm_A(\zb)$ are
bosonized using right-moving bosons $\phit^i_A(\zb)$.  In  terms
of bosons, the $R$ current is
\begin{align}
 J^3_R(z)  ={i\over 2} \sum_{A=1}^N(\partial\phi^5_A-\partial\phi^6_A)(z),\qquad
 J^\pm_R(z)=\sum_{A=1}^N e^{\pm i(\phi^5_A-\phi^6_A)(z)}.\label{R-current}
\end{align}
Note that $\Psi^s_A(z)$, $s=\pm$ have $R$-charge $J^3_R={s\over 2}$,
while $\Psit^\st_A(\zb)$, $\st=\pm$ have $\Jt^3_R={\st\over 2}$.
%These satisfy OPE
%\begin{align}
% J^3_R(z_1)J^3_R(z_2)&\sim {N\over 2(z_1-z_2)^2},\qquad
% J^3_R(z_1)J^{\pm}_R(z_2)\sim \pm{N\over z_1-z_2} J^{\pm}_R(z_2).
%\end{align}

The charge associated with the global $\widetilde {SU(2)}_I$ symmetry is
given by
\begin{align}
  \It^i&=\It^i_{\rm hol}+\It^i_{\rm antihol},
 \label{I-current}
\end{align}
where
\begin{align}
  \It^i_{\rm hol}&=\int{dz\over 2\pi i}\,\It^i_{\rm hol}(z),\qquad
 \It^i_{\rm antihol}=\int{d\zb\over 2\pi i}\,\It^i_{\rm antihol}(\zb).
\end{align}
The currents $\It^i_{\rm hol}(z)$, $\It^i_{\rm antihol}(\zb)$ are given
by
\begin{equation}
\begin{split}
 \It_{\rm hol}^3(z) &=
 {i\over 2}\sum_{A=1}^N(\partial\phi^5_A+\partial\phi^6_A)(z),
 \qquad
 \It_{\rm hol}^\pm(z)=
 \sum_{A=1}^N e^{\pm i(\phi^5_A+\phi^6_A)(z)},
 \\
 \It_{\rm antihol}^3(\zb)&=
 {i\over 2}\sum_{A=1}^N(\partialb\phit^5_A+\partialb\phit^6_A)(\zb),
 \qquad
 \It_{\rm antihol}^\pm(\zb)=
 \sum_{A=1}^N e^{\pm i(\phit^5_A+\phit^6_A)(\zb)},
\end{split}
\end{equation}
where we omitted the part that contains $x^a$ fields only, which is not
relevant for us; see \cite{David:2002wn} for the complete expression.
The charge $I^i$ associated with the global $SU(2)_I$ does not involve
fermions and can also be found in see \cite{David:2002wn}.  Note that
holomorphic part $\It^i_{\rm hol}$ and the antiholomorphic part
$\It^i_{\rm antihol}$ are not separately conserved; the chirality of the
CFT fields is not aligned with the chirality of the $SO(4)_I\cong
SU(2)_I\times\widetilde{SU(2)}_I$ symmetry.

It turns out
to be convenient to define $\Phi^\alphat_A(z)$ by
\begin{equation}
 \begin{split}
 \Phi^+_A(z)&=\Psi^+_A(z)={1\over \sqrt{2}}(\psi^1_A+i\psi^2_A)=e^{ i\phi^5_A(z)},\\
 \Phi^-_A(z)&=\Psi^-_A(z)^\dagger ={1\over \sqrt{2}}(\psi^3_A-i\psi^4_A)=e^{-i\phi^6_A(z)}.
\end{split}\label{def_Phi}
\end{equation}
We similarly define $\Phit_A^\alphat(\zb)$.  Note that
$\Phi^\alphat_A(z)$ and $\Phit^\alphat_A(\zb)$ have $\It$-charge
$\It^3={\alphat\over 2}$.

In the $S_N$ orbifold CFT \eqref{S_orb}, there are twist fields
$\sigma_P^{}(z)$, $P\in S_N$, which permute the copies of CFT as
$1\to P(1)$, $2\to P(2)$, \dots, $N\to P(N)$ as one circles the
point of insertion of $\sigma_P^{}$ \cite{Dixon:1986qv}.  For
example, if we have $\sigma_{(12\dots n)}^{}(z)$ at $z=0$, we
should impose boundary condition on the fields $x_A(z),\psi_A(z)$
as follows:
\begin{equation}\label{twist_bc}
 \begin{split}
  x_1(e^{2\pi i}z)&=x_{2}(z),~  \dots,~  x_n(e^{2\pi i}z)=x_{1}(z),\\
  \psi_1(e^{2\pi i}z)&=\pm\psi_{2}(z),~  \dots,~  \psi_n(e^{2\pi i}z)=\pm\psi_{1}(z),
\end{split}
\end{equation}
where ``$+$'' is for the NS sector and ``$-$'' is for the R sector.
%We
%can define
%\begin{align}
% x(ze^{2\pi i(A-1)})=x_A(z),\qquad
% \psi(ze^{2\pi i(A-1)})=(\pm 1)^{A-1}\psi_A(z),\qquad A=1,\dots,n
%\end{align}
%
%For example, if one takes an $n$-cycle
%$P=(12\dots n)$ and inserts $\sigma_{(12\cdots n)}^{}$ at $z=z_0$, the
%copies $A=1,2,\dots,n$ of CFT permute into each other as $1\to
%2\to\cdots\to n\to 1$ as one circles $z=z_0$.
This permutation of CFTs can be conveniently realized by going to a
covering space on which the fields of the CFT are single-valued
\cite{Lunin:2000yv, Lunin:2001pw}.  In the case of $\sigma_{(12\cdots
n)}^{}(z=0)$, one can define a new coordinate $t$ by
\begin{align}
 t^n = b z \qquad \text{near $z=0$},\label{t-spc}
\end{align}
so that circling $n$ times around $z=0$ corresponds to circling around
$t=0$ once.  This corresponds to inserting a twist field at $z=0$ in the
$z$-space that has the lowest conformal weight $\Delta_n={1\over
4}(n-{1\over n})$ \cite{Lunin:2000yv, Lunin:2001pw}.  We will
denote this twist operator henceforth by $\sigma_n(z)$.  Twist fields
with higher conformal weight are obtained by inserting some fields at
$t=0$ in the $t$-space.
In this way, computing correlation function of twist fields reduces to
finding a holomorphic map between the $z$-space and the covering
$t$-space that realizes the twists \cite{Lunin:2000yv, Lunin:2001pw}.
Note that this method of covering space is applicable only to $S_N$
orbifolds, and not applicable to general non-abelian orbifolds
\cite{Dijkgraaf:1989hb}.

One example of the operators that can be obtained by inserting a field
at $t=0$ is the chiral primary operator $\sigma^{--}_{n}(z)$ in the NS
sector.  Concretely, $\sigma_n^{--}(z)$ is obtained by inserting
in the $t$-space the following operator \cite{Lunin:2001pw}:
\begin{align}
 \sigma^{--}_n(t)\equiv \sigma^{-}_n(t)\,\sigmat^{-}_n(\tb),\label{sigma_n^--}
\end{align}
where
\begin{align}
 \sigma^{-}_n(t)\equiv b^{-p^2/n}e^{ip(\phi^5-\phi^6)}(t),\qquad
 \sigmat^{-}_n(\tb)\equiv \bar b^{-p^2/n}e^{ip(\phit^5-\phit^6)}(\tb).\label{sigma_n^-}
\end{align}
Here $p\equiv {n-1\over 2}$, and $\phi^i(t),\phit^i(\tb)$ are the lift
of $\phi^i_A(z),\phit^i_A(\zb)$ to the $t$-space, whose OPE is
$\phi^i(t_1)\phi^j(t_2)\sim -\log(t_1-t_2)$.  Note that
$\sigma^{--}_n(z)$ is not the coordinate transformation of
$\sigma^{--}_n(t)$, but it is a ``product'' of the pure twist operator
$\sigma_n(z)$ and the insertion $\sigma^{--}_n(t)$. Therefore, the
conformal dimension of $\sigma^{--}_n(z)$ is given by
\begin{align}
 h&=\Delta_n+{1\over n} \left({p^2\over 2}+{p^2\over 2}\right)
 = {n-1\over 2}.
\end{align}
Here $\Delta_n={1\over 4}\left(n-{1\over n}\right)$ is the conformal
dimension of the pure twist $\sigma_n(z)$.  In the second term, we
divided the conformal dimension in the $t$-space by $n$ to obtain the
conformal dimension in the $z$-space (remember that $z\propto t^n$)
\cite{Lunin:2001pw}. Similarly one can show that $(h,\tilde
h)=(j_R^3,\jt_R^3)=({n-1\over 2},{n-1\over 2})$. The chiral primary
$\sigma_n^{--}(z)$ has the smallest conformal dimension among the chiral
primary operators constructed on $\sigma_n(z)$ \cite{Lunin:2001pw}.
Another important fact is that $\sigma_1^{--}(z)$ is nothing but the
unit operator.

%\footnote{In Ref.\ \cite{Lunin:2001pw}, the
%twist operator $\Sigma^{(n-1)/ 2}_{(12\dots n)}$ is called
%$\sigma^{--}_n$.  See Table \ref{chprNSR} for the relation between the
%notation in this paper and the one in \cite{Lunin:2001pw}.}

%Using this method of covering space, one can construct a chiral primary
%operator $\sigma^{--}_{n}$ with $h=j_R^3=\tilde
%h=\jt_R^3={n-1\over 2}$ that has the smallest conformal weight for given
%$n$ \cite{Lunin:2001pw}.
%%\footnote{In Ref.\ \cite{Lunin:2001pw}, the
%%twist operator $\Sigma^{(n-1)/ 2}_{(12\dots n)}$ is called
%%$\sigma^{--}_n$.  See Table \ref{chprNSR} for the relation between the
%%notation in this paper and the one in \cite{Lunin:2001pw}.}
%Here,
%$n=1,2,\dots, N$\@.  $n=1$ corresponds to unit operator.  Concretely,
%the operator $\sigma^{--}_{n}$ is a product of the twist
%operator $\sigma_{n}$ which simply permutes the CFTs and a
%non-twist operator which contains fermion fields $\Psi^a_A$ only.

The twist fields $\sigma_P^{}(z)$, $P\in S_N$, considered above are not
proper fields of the orbifold CFT\@.  A proper field of the $S_N$
orbifold CFT should be invariant under conjugation by any element of
$S_N$.  This means that the twist sector is in one-to-one correspondence
with the conjugacy class of $S_N$ \cite{Dijkgraaf:1989hb}.  One can
construct a proper field from $\sigma_P^{}(z)$ by
\begin{align}
 \Sigma_P(z)={\lambda_P\over N!}\sum_{Q\in S_N}\sigma_{QPQ^{-1}}^{}(z),\label{mk_proper_twist}
\end{align}
where $\lambda_P$ is a normalization constant.  As long as we do
this summation over $S_N$ at the end of the computation of
correlation function, we can consider the cyclic permutation
$\sigma^{}_{(12\dots n)}(z)\equiv\sigma_n(z)$, instead of
$\sigma_P(z)$ with general $P\in S_N$.
%For example, the proper
%chiral primary field with an $n$-cycle is
%\begin{align}
% \Sigma^{--}_n&\equiv {\lambda_n\over N!}
% \sum_{Q\in S_N}\sigma^{--}_{Q(12\dots n)Q^{-1}}~.\label{Sigma_def}
%\end{align}

%
%The conjugacy class of $S_N$ is in one-to-one correspondence with
%partition of $N$:
%\begin{align}
% \sum_{n=1}^N N_n n= N, \qquad k_n=0,1,2,\dots.
%\end{align}
%Therefore, the twist sector of the orbifold CFT can be specified by the
%partition $\{N_n\}$. $N_n$ is the number of $n$-cycles.
%

Let us mention here one important aspect of the covering space
method. As we discussed above, computing correlation functions of twist
fields reduces to the problem of finding a holomorphic map $z=f(t)$ that
realizes the twists.  This coordinate transformation leads to a
nontrivial Liouville action $S_L[f(t)]$, which contributes to the
correlation function as $e^{S_L[f(t)]}$ \cite{Lunin:2000yv,
Lunin:2001pw}.  This Liouville factor is important when, for example,
computing the correlation function of twist fields at $z=z_i\,$; the
precise form of the map $f(t)$ depends on $z_i$, and this in turn leads
to a nontrivial dependence of the correlation function on
$z_i-z_j$. Actually the normalization in \eqref{sigma_n^-}, which
depends also on the map $f(t)$, also gives a nontrivial
contribution. However, what we will be interested in in this paper is
the $z_i$ dependence of the correlation functions of {\em non-twist\/}
operators at general points $z=z_i$ and twist operators at fixed points
$z=0,\infty$.  Because $e^{S_L}$ and the normalization of twist
operators depend only on the coordinates of twist operators, they are
irrelevant for us and we will ignore them altogether.

\subsection{Chiral primaries and spectral flow to R sector}\label{app:cft:Rgnd}

We are interested in the R sector ground states of the D1-D5
CFT\@. The R ground states can be obtained by first finding chiral
primary operators in the NS sector, and then spectral flowing to
the R sector.

%%%
%%%
\begin{table}[tbp]
 \begin{align*}
 \footnotesize
 \renewcommand{\arraystretch}{1.5}
 \begin{array}{|@{~~}c@{\,}c@{\,}l|c|c|c|c|}
 \hline
  \multicolumn{3}{|c|}{\text{chiral primaries}}
  & \!H^{2h,2\tilde h}(B)\!
% &\!\!{(h,\tilde h)_{\rm NS}~~\atop~~=(j^3_R,\jt^3_R)_{\rm NS}^{}}\!\!
  &\!\!{(h,\tilde h)_{\rm NS}=(j^3_R,\jt^3_R)_{\rm NS}^{}}\!\!
%  &\!\!\!\!{SU(2)_I~~\atop~~\times \widetilde{SU(2)}_{I}}\!\!\!\!
  &\!\!{SU(2)_I\times \widetilde{SU(2)}_{I}}\!\!
  &\!\!(j^3_R,\jt^3_R)_{\rm R}^{}\!\!
  \\[.75ex]
 \hline
 \hline
  \sigma_n^{--}&&
  & H^{0,0}
      & ({n-1\over 2},{n-1\over 2})
      &\bf (1,1)
          &(-\half,-\half)
              \\
  \hline\hline
  \tau_n^{\alphat,\st=-}&=&\Phi^\alphat_A\,\sigma_n^{--}
  & H^{1,0}
      & ({n\over 2},{n-1\over 2})
          &\bf (1,2)
      &(0,-\half)
              \\
  \hline
  \tau_n^{s=-,\alphat}&=&\Phit^\alphat_A\,\sigma_n^{--}
  & H^{0,1}
      & ({n-1\over 2},{n\over 2})
          &\bf (1,2)
      &(-\half,0)
              \\
  \hline\hline
  \sigma_n^{+-}&=&\Phi^+_A\Phi^-_A\,\sigma_n^{--}
  & H^{2,0}
      & ({n+1\over 2},{n-1\over 2})
          &\bf (1,1)
      &(\half,-\half)
              \\
 \hline
  \sigma_n^{\alphat\betat}&=&\Phi^\alphat_A\Phit^\betat_A\,\sigma_n^{--}
  & H^{1,1}
      & ({n\over 2},{n\over 2})
          &\bf (1,3)\oplus (1,1)
      &(0,0)
              \\
  \hline
  \sigma_n^{-+}&=&\Phit^+_A\Phit^-_A\,\sigma_n^{--}
  & H^{0,2}
      & ({n-1\over 2},{n+1\over 2})
          &\bf (1,1)
      &(-\half,\half)
              \\
 \hline\hline
  \tau_n^{s=+,\alphat} &=& \Phi^+_A\Phi^-_A \Phit^\alphat_A\,\sigma_n^{--}
  & H^{2,1}
      & ({n+1\over 2},{n\over 2})
          &\bf (1,2)
      &(\half,0)
              \\
 \hline
  \tau_n^{\alphat,\st=+}&=&\Phi^\alphat_A\Phit^+_A\Phit^-_A\,\sigma_n^{--}
  & H^{1,2}
      & ({n\over 2},{n+1\over 2})
          &\bf (1,2)
      &(0,\half)
              \\
 \hline\hline
  \sigma_n^{++}&=&\Phi^+_A\Phi^-_A\Phit^+_A\Phit^-_A\,\sigma_n^{--}
  & H^{2,2}
      & ({n+1\over 2},{n+1\over 2})
          &\bf (1,1)
      &(\half,\half)
              \\
 \hline
 \end{array}
\end{align*}
 \caption{\sl Single-trace chiral primaries in the NS sector.
 Here $n=1,2,\dots,N$, and $\alphat,\betat=\pm$.  Summation over
 $A=1,2,\dots, N$ is implied in all expressions, even if $A$ appears only
 once ({\it i.e.}, $\Phi_A^\alphat=\sum_{A=1}^N \Phi_A^\alphat$).  Note
 that $\sigma_1^{--}=1$.
 The corresponding cohomology $H^{2h,2\tilde h}(B)$, and the weights
 and charge of the chiral primary field are shown.  The $R$-charge of
 the corresponding R ground state is also shown.
 %
% For expressions in terms of bosons $\phi^{5,6}$, see text.
 } \label{chprNSR}
\end{table}
%%%
%%%

So, let us first focus on the chiral primaries of the orbifold CFT
\eqref{S_orb} in the NS sector.
%
%The chiral primaries with weight $(h,\tilde h)$ of $\CN=(4,4)$ SCFT on a
%manifold $K$ correspond to the elements of the cohomology
%$\CH_{2h,2\tilde h}(K)$ \cite{Witten:1982df}.  Therefore, chiral
%primaries can be constructed by taking the product of the chiral
%primaries corresponding to the cohomology of $B$ (the diagonal $T^4$,
%{\it i.e.}, the sum of all copies of $T^4$) and the chiral primary
%$\sigma_n^{--}(z)$ defined in \eqref{sigma_n^--}.
%
The chiral primaries with weight $(h,\tilde h)$ of $\CN=(4,4)$ SCFT on a
manifold $K$ correspond to the elements of the cohomology $H^{2h,2\tilde
h}(K)$ \cite{Witten:1982df}.  In the present case of the orbifold
$K=(T^4)^N/S_N$, the cohomology $H^*(K)$ can be constructed as follows
\cite{Vafa:1994tf}.  Let the basis of $H^*(B)$ be $w^a$,
$a=1,2,\dots,{\rm dim}(H^*(B))=16$, where $B$ is the diagonal $T^4$,
{\it i.e.}, the sum of all copies of $T^4$.  For each $w^a$, introduce a
``1-particle creation operator'' $\alpha_{-n}^a$, $n=1,2,\dots$.  Then
there is a one-to-one correspondence between the elements of $H^*(K)$
and the states in the ``Fock space'' generated by $\alpha_{-n}^a$.
Namely, for each element of $H^*(K)$, there is a state $\prod_{n,a}
({\alpha_{-n}^a})^{N_{na}}\ket{0}$, $\sum_n nN_n=N$.  If $w^a$ is an
even (odd) form, $\alpha_{-n}^a$ is bosonic (fermionic).

In the present case, $H^*(B)$ has 8 elements of even rank and 8 elements
of odd rank, so there are corresponding 8+8 chiral primaries.  The
chiral primaries corresponding to $\alpha_{-n}^a$ is constructed by
multiplying the above 8+8 chiral primaries with the chiral primary
$\sigma_n^{--}(z)$ defined in \eqref{sigma_n^--}, which is a twist
operator of order $n$.
In Table \ref{chprNSR}, we list all chiral primary fields in the NS
sector that correspond to the ``1-particle creation operator''
$\alpha_{-n}^a$.  They are {\em single-trace\/} in the sense that they
involve only one summation over copies $\sum_{A=1}^N$ and only one twist
operator $\sigma_n^{--}(z)$.
We also present their conformal weight, $R$-charges, and $SU(2)_I\times
\widetilde{SU(2)}_I$ charges, as well as the $R$-charges of the R ground
states that can be obtained by spectral flow, using
\eqref{spectral_flow}.
One sees that there are 8 bosonic and 8 fermionic single-trace chiral
primaries:
\begin{align}
 \sigma_n^{s\st},~ \sigma_n^{\alphat\betat}, ~ \tau_n^{s\alphat}, ~\tau_n^{\alphat\st}.
\end{align}
Here, $s,\st=\pm$ correspond to $SU(2)_R\times \widetilde{SU(2)}_R$
charges $(J_R^3,\Jt_R^3)=({s\over 2},{\st\over 2})$, while
$\alphat,\betat=\pm$ correspond to $\widetilde{SU(2)}_I$ charge
$\It^3={\alphat\over 2}$.  The fields
$\tau_n^{s\alphat},\tau_n^{\alphat\st}$ which correspond to odd-rank
elements of $H^*(B)$ are indeed fermionic because $\Phi$'s
anticommute. These single-trace chiral primaries are known to be in
one-to-one correspondence with the Kaluza--Klein spectrum of
particle supergravity on $AdS_3\times S^3\times T^4$ \cite{deBoer:1998ip}.  We
will use $\sigma^{--}_n$, $\sigma^{++}_n$, etc.\ with explicit $+$, $-$
signs exclusively for denoting the $\sigma^{s\st}_n$ operators, not
$\sigma^{\alphat\betat}_n$.

Now let us consider spectral flowing to the R sector.  From
\eqref{R-current}, the spectral flow operator that maps NS sector
operators to the R sector operators is
\begin{align}
 U(z)&=\exp\left[-{i\over 2}\sum_{A=1}^N (\phi_A^5-\phi_A^6)\right](z),\label{spctrl_flow(z)}
\end{align}
where we wrote the holomorphic part only.  The spectral flow relates the
charges in the R and NS sectors as follows:
\begin{align}
 h_{\rm R}&=h_{\rm NS}-(j^3_R)_{\rm NS}^{}+{c\over 24},
 \qquad
 (j_R^3)_{\rm R}^{}=(j_R^3)_{\rm NS}^{}-{c\over 12}.
\label{spectral_flow}
\end{align}
Here, roman R stands for Ramond, while italic $R$ is for
$R$-charge.
The spectral flow operator in the $t$-space is given by coordinate
transformation of $U(z)$ by:
\begin{align}
 U(t=0)&\propto \exp\left[-{i}{n\over 2}(\phi^5-\phi^6)\right](t=0),\label{spctrl_flow(t)}
\end{align}
where we used $\phi_A(z)\xrightarrow[]{} \phi(te^{2\pi i
(A-1)/n})\xrightarrow[t\to 0]{}\phi(t=0)$. For example, we can use this
to map the NS sector twist operator $[\sigma^{-}_n(t)]_{\rm NS}\propto
\exp[i{n- 1\over 2}(\phi^5-\phi^6)](t)$ into the R sector:
\begin{align}
 [\sigma^{-}_n(t)]_{\rm NS} \to
 [\sigma^{-}_n(t)]_{\rm R}\propto
 e^{- {i\over 2}(\phi^5-\phi^6)}(t).
 \label{sigma_R}
\end{align}
We can check that this has the correct conformal dimension and
$R$-charge as a R ground state:
\begin{align}
 h_{\rm R}&=\Delta_n+{1\over n}\left[\half\left(\half\right)^2+\half\left(\half\right)^2\right]
 ={n\over 4}={c\over 24},
 \qquad
 (j_R^3)_{\rm R}=- \half.
\end{align}
Including other operators, the list of R  ground states
corresponding to single-trace NS chiral primaries is
\begin{equation}
 \begin{split}
 [\sigma_n^{s\st}(t)]_{\rm R}&=
 e^{{i s\over 2}(\phi^5-\phi^6)}(t)\, e^{{i \st\over 2}(\phit^5-\phit^6)}(\tb),\qquad
 [\sigma_n^{\alphat\betat}(t)]_{\rm R}=
 e^{{i \alphat\over 2}(\phi^5+\phi^6)}(t)\, e^{{i\betat\over 2}(\phit^5+\phit^6)}(\tb),
 \\
 [\tau_n^{s\alphat}(t)]_{\rm R}&=
 e^{{i s\over 2}(\phi^5-\phi^6)}(t)\, e^{{i \alphat\over 2}(\phit^5+\phit^6)}(\tb),\qquad
 [\tau_n^{\alphat\st}(t)]_{\rm R}=
 e^{{i \alphat\over 2}(\phi^5+\phi^6)}(t)\, e^{{i \st\over 2}(\phit^5-\phit^6)}(\tb).
\end{split}\label{sngl-tr_RR_gnd}
\end{equation}
Here, we ignored normalization constants because they are
irrelevant for our purposes as explained at the end of the last
subsection.  We will call these operators \eqref{sngl-tr_RR_gnd}
single-trace R ground states.  Henceforth, we restrict ourselves
to the R sector and drop the subscript R from the twist operators
\eqref{sngl-tr_RR_gnd}.

General R ground states are obtained by multiplying the single-trace R
ground states \eqref{sngl-tr_RR_gnd} together.  They can be written as
\begin{equation}
\begin{split}
 \sigma&= \prod_{n,\mu} (\sigma_{n}^{\mu})^{N_{n\mu}} (\tau_{n}^\mu)^{N'_{n\mu}},
 \\
 \sum_{n,\mu}n (N_{n\mu}+N'_{n\mu})&=N, \qquad
 N_{n\mu}=0,1,2,\dots,\quad N'_{n\mu}=0,1,
\end{split}\label{gen_twist}
\end{equation}
where $\mu$ labels the 8 polarizations of bosons and fermions, {\it
i.e.}, $\mu=(s,\st),(\alphat,\betat)$ for bosons
$\mu=(s,\alphat),(\alphat,\st)$ for fermions.  The numbers
\begin{align}
 \{N_{n\mu},N_{n\mu}'\}&
\end{align}
uniquely specify the R ground state.  We will refer to the factors
$\sigma^\mu_n,\tau^\mu_n$ in \eqref{gen_twist} as constituent twist
operators of the twist operator $\sigma$.

\subsection{Correlation function of non-twist operators}\label{app:cft:corrfunc}

We would now like to compute correlation functions in the R ground
states of the D1-D5 CFT\@.  Such CFT correlation functions are related
to supergravity amplitudes in the dual geometry via AdS/CFT\@.

We want to compute the 2-point function of the ``probe'' operator
$\CA$ in the state created by a general twist operator.  Let us
assume that $\CA$ {\em does not contain twists\/} and can be written
as a sum over copies of the CFT:
\begin{align}
 \CA&={1\over \sqrt{N}}\sum_{A=1}^N \CA_A, \label{sum_CA_A}
\end{align}
where $\CA_A$ is a {\em non-twist\/} operator that lives  in the
$A$-th copy.  
For example, we can take
\begin{align}
 \CA_A=\partial X^a_A(z) \bar \partial X^b_A(\zb),
 \label{A_grvtn}
\end{align}
which corresponds to fluctuation of the metric in the internal $T^4$
direction.  
%Such non-twist operators are only a subset of the
%operators that correspond to the excitations in the bulk; there are
%bulk modes that correspond to twist operators also.  We will
%restrict ourselves to non-twist operators in this paper because
%their correlation functions are much easier to compute than those of
%twist operators.  We  will see that the correlation function of
%non-twist operators is enough to see that an ``effective'' geometry
%emerges in the $N\to \infty$ limit.

Let us consider the general R ground state \eqref{gen_twist}.  If
we denote $\sigma_n^\mu,\tau_n^\mu$ collectively by
$\sigma_n^{\muh}$, and $N_{n\mu},N_{n\mu}'$ by $N_{n\muh}$, then
we can write \eqref{gen_twist} as
\begin{align}
 \sigma&= \prod_{n,\muh} (\sigma_n^{\muh})^{N_{n\muh}}.
\end{align}
The correlation function of the probe operator $\CA$ in this state is,
taking into account the summation over copies (Eq.\ \eqref{mk_proper_twist}),
\begin{align}
 \bracket{\Sigma^\dagger\CA^\dagger  \CA \Sigma}
 &=\bracket{\sigma^\dagger \CA^\dagger \CA \sigma}
 ={1\over N}\sum_{A,B=1}^N \bracket{\sigma^\dagger \CA_A^\dagger
  \CA_B \sigma}\notag\\
 &={1\over N}\sum_{A,B=1}^N \Bracket{
   \Bigl[\prod_{n,\muh} (\sigma_{n}^{\muh})^{N_{n\muh}}\Bigr]^\dagger
   \CA_A^\dagger \CA_B \Bigl[\prod_{\nu,\muh}
   (\sigma_{n}^\muh)^{N_{n\muh}}\Bigr]}
 \notag\\
 &={1\over N}\sum_{n,\muh}
 N_{n\muh} \sum_{A,B\in \sigma_n^\muh}
 \bracket{ [\sigma_{n}^\muh]^\dagger \CA_A^\dagger
 \CA_B  \sigma_{n}^\muh }\notag\\
 &={1\over N}\sum_{n,\muh} n N_{n\muh} \sum_{A=1}^{n}
 \bracket{ [\sigma_{(1\cdots n)}^{\muh}]^\dagger \CA_A^\dagger \CA_1
   \sigma_{(1\cdots n)}^{\muh} },
 \label{SAAS}
\end{align}
where $\sum_{A,B\in \sigma_n^\muh}$ means to sum over copies $A,B$ that
are involved in the $n$-cycle of $\sigma_{n}^\muh$.  In the first
equality, we used the fact that $\CA$ is a sum over copies,
\eqref{sum_CA_A}.  In the fourth equality, we used the fact that the
``initial'' and ``final'' states must have the same length of twist and
the same $SU(2)$ charges to give a nonvanishing correlator, since
$\CA_A^\dagger \CA_B$ does not involve twist or charges.  We assumed
that the three point function vanishes:
$\bracket{[\sigma_n^\muh]^\dagger\CA_A \sigma_n^\muh}=0$, which is true
in the case considered in this paper.
Note that the final expression \eqref{SAAS}  decomposed into
contributions from constituent twist operators.  This is because
we are restricting ourselves to non-twist probes $\CA$, and
because we are in the orbifold point approximation and ignoring
interactions.  Once we start considering twist probes or
interaction, this will no longer be the case.

Therefore, for a non-twist operator $\CA$ in the orbifold approximation,
all we have to compute is the 4-point function
\begin{align}
 \Bracket{ [\sigma_{(1\cdots n)}^{\muh}(z=\infty)]^\dagger \CA_A(z_1)^\dagger
 \CA_B(z_2) \sigma_{(1\cdots n)}^\muh(z=0)}
 \equiv
 \bracket{ \CA_A(z_1)^\dagger
 \CA_B(z_2) }_{\sigma_{(1\cdots n)}^\muh},
 \label{4pt_func}
\end{align}
where $1\le A,B\le n$.

\subsection{Boson correlation function}\label{app:cft:bosoncorr}

Let us evaluate the correlation function \eqref{SAAS}, \eqref{4pt_func}
for $\CA$ a purely bosonic non-twist operator such as $\partial
X\bar\partial X$.  In this case, we can replace the twist operator
$\sigma_{(1\dots n)}^{\muh}(z)$ in \eqref{4pt_func} with the pure twist
operator $\sigma_{(1\dots n)}(z)$, since these two are different only in
their fermionic dressing, to which the bosonic operator $\CA$ is
insensitive.

The twist operators $\sigma_{(1\dots n)}(z)$ at $z=0,\infty$ mean
that $X_A^a(z)$ permute as $X_1^a\to X_2^a\to\cdots\to X_n^a\to
X_1^a$ as one circles $z=0,\infty$.  As explained around
\eqref{t-spc}, we can conveniently go to the covering $t$-space by
\begin{align}
 z=bt^n\label{z=bt^n}
\end{align}
on which we have  single-valued fields $X^a(t)$.  If we normalize the
correlation function in the $t$-space as
\begin{align}
 \bracket{\CA_A^\dagger(t_1)\CA_B(t_2)} = {C \over (t_1 -t_2)^{2h}(\tb_1 -\tb_2)^{2\tilde h}},
\end{align}
where $(h,\tilde h)$ is the conformal weight of $\CA$, then the
correlation function on the $z$-plane is
\begin{align}
 \bracket{\CA_A^\dagger(z_1) \CA_B(z_2)}_{\sigma_{(1\cdots n)}}
 =
 {C\over n^{2h+2\tilde h}(z_1 z_2)^h(\zb_1 \zb_2)^{\tilde h}
  \Bigl[\bigl({z_1\over z_2}\bigr)^{1\over 2n}
     -\bigl({z_2\over z_1}\bigr)^{1\over 2n}\Bigr]^{2h}
  \Bigl[\bigl({\zb_1\over \zb_2}\bigr)^{1\over 2n}
     -\bigl({\zb_2\over \zb_1}\bigr)^{1\over 2n}\Bigr]^{2\tilde h}
  }.
 \notag
\end{align}
If we go to the cylinder coordinate $w$ by
\begin{align}
 z=e^{-iw},
\end{align}
then the correlation function is
\begin{align}
 \bracket{\CA_A^\dagger(w_1) \CA_B(w_2)}_{\sigma_{(1\cdots n)}}=
 {C \over \left[2 n \sin\left({w \over 2n }\right)\right]^{2h}
 \left[2n \sin\left({\wb \over 2n}\right)\right]^{2\tilde h}},
 \label{corrAA_w}
\end{align}
where
\begin{align}
 w &\equiv w_1-w_2,\qquad
 \wb \equiv \wb_1-\wb_2.
\end{align}
Here, the copy labels $A,B$ mean that $w_1$ and $w_2$ must  be
understood as $w_1+2\pi (A-1)$ and $w_2+2\pi (B-1)$, respectively.

The result (\ref{corrAA_w}) expresses the fact that the effective
circumference of the CFT cylinder is $2\pi n$, where the factor of $n$
comes from the permutation of $n$ copies of the CFT\@.  Indeed, from
this picture one can easily write down (\ref{corrAA_w}) directly, simply
by inserting the appropriate factors of $n$ in the usual free correlator
on the cylinder.

\subsection{Fermion correlation function}\label{app:cft:fermicorr}

Now let us evaluate the correlation function \eqref{4pt_func} in the
case where $\CA$ involves fermions $\psi$.  As an example, let us
consider
\begin{align}
 G_z
 &\equiv
 \bracket{\sigma_n^{--}(z_\infty)^\dagger
 \Psi_A^+(z_1)^\dagger \Psi_B^+(z_2)
 \sigma_n^{--}(z_0)}.
\end{align}
It is understood that we will take $z_\infty\to\infty,z_0\to 0$ in the
end, so we can use \eqref{z=bt^n} as the relation between $z$ and $t$
coordinates.  Using the expression of operators in terms of bosons
(Eqs.\ \eqref{bosonize}, \eqref{sngl-tr_RR_gnd}), one computes
\begin{align}
 G_z&\propto
 e^{S_L}\bracket{
 [e^{{i\over 2}(\phi^5-\phi^6)}(t_\infty)e^{-{i\over 2}(\phit^5-\phit^6)}(\tb_\infty)]
 \, e^{-i\phi^5}(t_1)
 \, e^{i\phi^5}(t_2)
 \,
 [e^{-{i\over 2}(\phi^5-\phi^6)}(t_0)e^{{i\over 2}(\phit^5-\phit^6)}(\tb_0)]
 }\notag\\
 &\qquad\qquad
 \times \left({dt_1\over dz_1}\right)^{1/2} \left({dt_2\over dz_2}\right)^{1/2}\notag\\
 &\propto
 (t_\infty-t_1)^{-1/2} (t_\infty-t_2)^{1/2}
 (t_\infty-t_0)^{-1/2} (\tb_\infty-\tb_0)^{-1/2}
 (t_1-t_2)^{-1}(t_1-t_0)^{1/2}(t_2-t_0)^{-1/2}\notag\\
 &\qquad\qquad
 \times
 \left({dt_1\over dz_1}\right)^{1/2} \left({dt_2\over dz_2}\right)^{1/2}\notag\\
 &\to
 t_\infty^{-1/2}\tb_\infty^{-1/2}
 (t_1-t_2)^{-1}t_1^{1/2}t_2^{-1/2}
 \left({dt_1\over dz_1}\right)^{1/2} \left({dt_2\over dz_2}\right)^{1/2}~,\qquad
 (t_\infty\to\infty,t_0\to 0)~.
\end{align}
$S_L$ is the Liouville action as explained at the end of subsection
\ref{app:cft:orb_cft}, which is an irrelevant factor for our purpose and
was dropped.  Now rewrite $t$ in terms of $z$ using $t\propto
z^{1/n}$. The factor $t_\infty^{-1/2}\tb_\infty^{-1/2}$, along with
the dropped Liouville factor $e^{S_L}$ and the normalization
constants of the twist operators, corresponds in the Lorentzian
signature simply to the phase $e^{-i E t}$ due to the initial and
final states $\sigma_n^{--}$.  Thus they are irrelevant and we will
drop this factor henceforth.  The result is
\begin{align}
 G_z=\bracket{\Psi_A^+(z_1)^\dagger \,\Psi_B^+(z_2)}_{\sigma_n^{--}}
 &\propto {1\over (z_1 z_2)^{1/2}[1-(z_2/z_1)^{1/n}]}.
\end{align}
Passing to the cylinder coordinate $w$ by $z=e^{-iw}$,
\begin{align}
 \bracket{\Psi_A^+(w_1)^\dagger \,\Psi_B^+(w_2)}_{\sigma_n^{--}}
 =\bracket{\Psi_A^+(z_1)^\dagger \,\Psi_B^+(z_2)}_{\sigma_n^{--}}
 \left({dz_1\over dw_1}\right)^{1/2}\left({dz_2\over dw_2}\right)^{1/2}
 &\propto {1\over 1-e^{iw/n}},
\end{align}
where  $w=w_1-w_2$.
As before, the copy labels $A,B$ mean that $w_1$, $w_2$ must be
understood as $w_1 + 2\pi (A-1)$, $w_2+ 2\pi(B-1)$,
respectively.  Therefore, more precisely,
\begin{align}
 \bracket{\Psi_A^+(w_1)^\dagger\, \Psi_B^+(w_2)}_{\sigma_n^{--}}
 &= {i\over n[1-e^{i(w+2\pi (A-B))/n}]}
 = {e^{i(w+2\pi(A-B))/2n}\over 2n\sin {w+2\pi(A-B)\over 2n}},
 \label{fermia}
\end{align}
where the normalization was fixed by requiring $\Psi_A^+(w_1)^\dagger\,
\Psi_B^+(w_2)\sim \delta_{AB}/w$.

Similarly, one can compute other correlators of the $SU(2)_R$ doublet
fields $\Psi^s_A$ as:
\begin{equation}
 \begin{split}
 \bracket{\Psi_A^{s'}(w_1)^\dagger \, \Psi_B^{s'}(w_2)}_{\sigma_n^{s\st}}
 &=
 {e^{iss'w/2n}\over 2n\sin{w\over 2n}},
 \quad
 \bracket{\Psi_A^{s'}(w_1)^\dagger \, \Psi_B^{s'}(w_2)}_{\sigma_n^{\alphat\betat}}
 =
 {e^{i\alphat w/2n}\over 2n\sin{w\over 2n}}
 \\
 \bracket{\Psi_A^{s'}(w_1)^\dagger \, \Psi_B^{s'}(w_2)}_{\tau_n^{\alphat\st}}
 &=
 {e^{i\alphat w/2n}\over 2n\sin{w\over 2n}},
 \quad
 \bracket{\Psi_A^{s'}(w_1)^\dagger \, \Psi_B^{s'}(w_2)}_{\tau_n^{s\alphat}}
 =
 {e^{i s\st w/2n}\over 2n\sin{w\over 2n}},
\end{split}
\label{ferm_corr_bldblk}
\end{equation}
where it is understood that $w$ really means $w+2\pi(A-B)$.   Or,
in terms of the $\widetilde{SU(2)}_I$ doublet fields $\Phi_A^\alphat$ defined
in \eqref{def_Phi},
\begin{equation}
 \begin{split}
 \bracket{\Phi_A^{\alphat}(w_1)^\dagger \, \Phi_B^{\alphat}(w_2)}_{\sigma_n^{s\st}}
 &=
 {e^{is w/2n}\over 2n\sin{w\over 2n}},
 \quad
 \bracket{\Phi_A^{\alphat}(w_1)^\dagger \, \Phi_B^{\alphat}(w_2)}_{\sigma_n^{\betat\gammat}}
 =
 {e^{i\alphat\betat w/2n}\over 2n\sin{w\over 2n}}
 \\
 \bracket{\Phi_A^{\alphat}(w_1)^\dagger \, \Phi_B^{\alphat}(w_2)}_{\tau_n^{\betat\st}}
 &=
 {e^{i\alphat\betat w/2n}\over 2n\sin{w\over 2n}},
 \quad
 \bracket{\Phi_A^{\alphat}(w_1)^\dagger \, \Phi_B^{\alphat}(w_2)}_{\tau_n^{s\betat}}
 =
 {e^{i s w/2n}\over 2n\sin{w\over 2n}}
\end{split}
\label{ferm_corr_bldblk2}
\end{equation}

Just as we remarked  after the derivation of the bosonic correlator
(\ref{corrAA_w}), these results for fermionic correlators express
the fact that the effective length of the CFT cylinder has increased
by a factor of $n$ due to the permutation, and the results could
have been obtained from this property alone.

%If we sum over the copies,
%\begin{align}
% \sum_{A=1}^n \bracket{\Psi_A^+(w_1)^\dagger\, \Psi_1^+(w_2)}_{\sigma_n^{--}}
% &= \sum_{k=0}^{n-1} {e^{-i(w+2\pi k)/2n}\over 2n\sin {w+2\pi k\over 2n}},
%\end{align}
%Note that this is periodic under $w\to w+2\pi$, as should be
%true for the R sector in the cylinder coordinate.
%Performing the summation using the formula
%\begin{align}
% \sum_{k=0}^{n-1}&{1\over a e^{2\pi i k/n}-1}={n\over a^n-1},
%\end{align}
%we obtain
%\begin{align}
% \sum_{A=1}^n \bracket{\Psi_A^+(w_1)^\dagger\, \Psi_1^+(w_2)}_{\sigma_n^{--}}
% &= { e^{-i w/2}\over 2\sin {w\over 2}}.
%\end{align}

\section{Statistical mechanics of the ensemble with $J\neq 0$}
\label{app:J=/=0}

In this appendix, we study the statistical mechanics of the ensemble
with $J\neq 0$ studied in subsection \ref{subsec:typ_J=/=0}.

For the special case $N_B=24$, $n_B=1$, $N_F=n_F=0$, {\it i.e.\/}\ for
the D1-D5 system on K3, state counting was first studied in
\cite{Russo:1994ev} from the heterotic dual viewpoint.  More recently,
statistical mechanics of the K3 case was analyzed in
\cite{Iizuka:2005uv}, and the following discussion is a more detailed
and generalized version of the one presented in \cite{Iizuka:2005uv}.

For the canonical ensemble of bosons and fermions with the spin
assignment \eqref{spin_assign}, the partition function is
\begin{align}
 Z(\beta,\mu)
% &=\sum_{N,J}d_{N,J}q^N z^J\label{fiuk27May05}\\
 &={\rm Tr}[e^{-\beta(N-\mu J)}]
 =\prod_{n=1}^\infty
 {[(1+z^{1/2}q^n)(1+z^{-1/2}q^n)]^{n_F}(1+q^n)^{N_F-2n_F}
 \over
 [(1-zq^n)(1-z^{-1}q^n)]^{n_B}(1-q^n)^{N_B-2n_B} }\notag\\
 &=2^{n_B-{N_F\over 2}}q^{N_B-N_F \over
 24}\eta(\tau)^{-N_B+3n_B-{N_F\over 2}}
 \left[{\vartheta_2({\nu\over 2}|\tau)\over\cos{\pi \nu\over 2}}\right]^{n_F}
 \left[{\sin\pi\nu\over \vartheta_1(\nu|\tau)}\right]^{n_B}
 \vartheta_2(0|\tau)^{{N_F\over 2}-n_F}.
\end{align}
Here $q=e^{2\pi i \tau}=e^{-\beta}$, $z=e^{2\pi i
\nu}=e^{\beta\mu}$.  After modular transformation, one obtains
the expression for $\beta\ll 1$:
%\begin{align}
% Z(\beta,\mu)&=2^{n_B-{n_F\over 2}}q^{N_B-N_F\over 24}(-i\tau)^{{N_B\over 2}-n_B}
% e^{{\pi i \nu^2\over \tau}(n_B-{n_F\over 4})}
% [\eta(-\tfrac{1}{\tau})]^{-N_B+3n_B-{N_F\over 2}}\notag\\
% &\qquad\times
% {(\sin\pi\nu)^{n_B}\over (\cos{\pi\nu\over 2})^{n_F}}
% {[\vartheta_4({\nu\over 2\tau}|-{1\over\tau})]^{n_F}
% [\vartheta_4(0|-{1\over\tau})]^{{N_F\over 2}-n_F}
% \over
% [i\vartheta_1({\nu\over \tau}|-{1\over\tau})]^{n_B}
% }.
%\end{align}
\begin{align}
 Z(\beta,\mu)
 &=
 2^{-{N_F\over 2}}
 e^{
 {\pi^2c\over 6\beta}-{\beta\over 24}(N_B-N_F)-{\mu^2\beta\over 2}
 (n_B-{n_F\over 4})
 }
 \left({\beta\over 2\pi}\right)^{{N_B\over 2}-n_B}
 \left[{\sinh{\beta\mu\over2}\over\sin\pi\mu}\right]^{n_B}
 {1\over[\cosh{\beta\mu\over 4}]^{n_F}}\label{jhws19Jul05}
\end{align}
up to exponentially suppressed terms by $e^{-{2\pi^2\over
\beta}}$. Here, $c\equiv N_B+N_F/2$.  For the ``Hamiltonian''
$N-\mu J$ to be positive definite, we must restrict the range of
the chemical potential to $|\mu|<1$.  Therefore, for $\beta\ll 1$,
one  can further simplify \eqref{jhws19Jul05} as
\begin{align}
 Z(\beta,\mu)\sim
% 2^{-{N_F\over 2}}\pi^{n_B}
 \beta^{N_B\over 2}
 \left({\mu\over\sin\pi\mu}\right)^{n_B}e^{\pi^2c/6\beta},\label{jjnv19Jul05}
\end{align}
up to a numerical factor.
Note that this result does not depend on $n_F$; all   spins are
carried by bosons.  This is because the Pauli exclusion principle
exacts a high price in $N$ when the fermions carry a macroscopic
amount of angular momentum.

Let us compute the entropy of this system by thermodynamic
approximation, {\it i.e.}, by saddle point approximation.
By the standard formula of thermodynamics,
\begin{align}
 N&=-\left({\partial\log Z\over \partial \beta}\right)_{\beta\mu}
 ={c\pi^2\over 6\beta^2}+{n_B\mu\over \beta}g(\mu),\qquad
 J=\left({\partial\log Z\over\partial (\beta\mu)}\right)_\beta
 ={n_B\over \beta}g(\mu),\label{jiym19Jul05}
\end{align}
where
\begin{align}
 g(\mu)&\equiv {1\over\mu}-{\pi\over\tan\pi\mu}
 ={\pi^2\mu\over 3}+{\pi^4\mu^3\over 45}+\cdots.\label{jjwf19Jul05}
\end{align}
In deriving \eqref{jiym19Jul05}, we ignored $\beta^{N_B/2}$ in
\eqref{jjnv19Jul05} in the thermodynamic approximation.  The entropy is
\begin{align}
 S=\beta(N-\mu J-F) =\beta(N-\mu J)+\log Z
 ={c\pi^2\over 3\beta}+n_B\log\left({\mu\over\sin\pi\mu}\right)
 .\label{jiyq19Jul05}
\end{align}
From \eqref{jiym19Jul05}, \eqref{jiyq19Jul05}, we obtain
\begin{align}
 S&=2\pi\sqrt{{c\over 6}(N-\mu J)}+n_B\log\left({\mu\over\sin\pi\mu}\right) ,\label{jdxt25Jul05}\\
 J&={3n_B^2\mu g(\mu)^2\over c\pi^2}\left[
 \sqrt{1+{2c N\pi^2\over 3\mu^2n_B^2g(\mu)^2}}-1 \right].\label{jjvo19Jul05}
\end{align}
From the relation \eqref{jjvo19Jul05} and the form of the function
$g(\mu)$ \eqref{jjwf19Jul05}, it is easy to see that $\mu\to {\rm
sign}(J)$ is needed in order that $|J|=\CO(N)$.  More precisely, we need
$|\mu-{\rm sign}(J)|\sim N^{-1/2}$.  Therefore,
\begin{align}
 S=\log d_{N,J}& \approx
 2\pi\sqrt{{c\over 6}(N-|J|)},
\end{align}
where we dropped the subleading log term coming from the second term in
\eqref{jdxt25Jul05}.  Note that, $\mu\to\pm 1$ implies that the energy
of bosons with $J=\pm 1$ vanishes and the Bose--Einstein condensation of
those bosons occurs.

\end{document}